\documentclass{imamat}

\usepackage{dcolumn,graphicx,amssymb,amsmath}
\usepackage[modulo,mathlines]{lineno}
\usepackage[usenames]{color}

\begin{document}

\title{Using electric fields to induce patterning in leaky dielectric fluids 
in a rod-annular geometry}

\shorttitle{Electrified layers} 
\shortauthorlist{Q. WANG AND D. T. PAPAGEORGIOU} 

\author{
\name{Qiming Wang$^*$}
\address{Department of Mathematics and Statistics, York University, Toronto, ON, M3J 1P3, Canada\email{$^*$Corresponding author: qmwang@yorku.ca}}
\name{Demetrios T. Papageorgiou}
\address{Department of Mathematics, Imperial College London, South Kensington Campus, London SW7 2AZ, UK}}

\maketitle

\begin{abstract}
{The stability and axisymmetric deformation of two immiscible, viscous, perfect or leaky dielectric fluids confined in the annulus between two concentric cylinders
are studied in the presence of radial electric fields. The fields are set up by imposing a constant voltage potential difference between the inner and outer cylinders.
We derive a set of equations for the interface in the long-wavelength approximation which retains the essential physics of the system and
allows for interfacial deformations to be as large as the annular gap hence accounting for possible touchdown at the inner
or outer electrode.
The effects of the electric parameters are evaluated
initially by performing a linear stability analysis which shows excellent agreement with the linear theory of the full axisymmetric problem in the appropriate long wavelength regime. 
The nonlinear interfacial dynamics are investigated by carrying out direct numerical simulations of the derived long wave models, both in the absence and presence
of electric fields. For non-electrified thin layer flows (i.e. one of the layers thin relative to the other) the
long-time dynamics agree with the lubrication approximation results found in literature. When the liquid layers have comparable
thickness
our results demonstrate the existence of both finite time and infinite time singularities (asymptotic touching solutions) in the system. 
It is shown that a two-side touching solution is possible for both the non-electrified and perfect dielectric cases, 
while only one-side touching is found in the case of leaky dielectric liquids, where the flattened interface shape resembles the pattern solutions found in literature. Meanwhile the finite-time singular solution agrees qualitatively with the experiments of~\cite{Reynolds1965}.
}
{electrohydrodynamics, interfacial instability, leaky dielectric fluids, rod annular flows, finite-time singularities}
\end{abstract}

\section{Introduction}\label{intro}
\setcounter{equation}{0}

A liquid layer coating the outside or inside of a cylindrical tube is subject to Rayleigh instability due to the action
the action of surface tension (for example see \cite{Goren} for a linear theory where two fluids can be present in what
is known as a quiescent core-annular flow - CAF). \cite{Hammond1983} derived a lubrication equation for thin films in the absence of gravity and a surrounding fluid; numerical and analytical solutions
suggest that the interface tends to touch the wall asymptotically as time increases - lobe and collar
structures form. More extended long-time dynamics of the Hammond equation
have been studied by~\cite{LRKCJ2006} who integrate the equations numerically to very
large times (and also provide multi scale asymptotic solutions) and show that collars and lobes can interact depending on the domain length. 
With gravity present (e.g. gravity-driven film flow coating the outside of a vertical cylinder) 
a similar equation has been investigated by ~\cite{KC94} and \cite{Kerchman95} who showed that the generation of shear-induced interfacial waves may suppress the instability. A weakly nonlinear version of the equation for the interface leads to the so called Kuramoto-Sivashinsky (KS) equation (see \cite{Frenkel87,Papageorgiou1990}). Other related studies can be found in \cite{KDB2001,CM2006,COO14} in slightly different asymptotic limits, with the Hammond equation emerging as an appropriate limit.
Core-annular flows of two immiscible fluids have been studied extensively and mostly in the weakly nonlinear regime 
or by direct numerical simulations - see, for example, \cite{Pozrikidis92jfm,BKJ96,LiRenardy99, KT2001,KT2002,BLP2006}. Unlike the case of single fluid flows,
the presence of shear introduces technical difficulties and prevents the asymptotically
rational derivation of fully nonlinear equations (with interfacial amplitudes scaling with the tube radius, for example),
making analytical progress difficult and restricted to the useful weakly nonlinear regime. 
Studies of strongly nonlinear interfacial deformations that take into account fluid-fluid coupling are therefore limited even in the
tractable case in the absence of background shear. However, see the long wave model in \cite{Wang2013} when the core fluid is much more viscous than the annulus (this is an extension of the single jet work by \cite{EgDu1994}), and an integral-boundary-layer model in \cite{DR15} when the viscosity ratio is of order one. 

When the two-fluid flow is confined in a cylindrical annulus and a shear is imposed by either a driving axial pressure gradient or
the axial motion of the inner or outer cylinder, the flow is known as rod-annular flow and has different physical characteristics from
usual core-annular flows when the inner rod is absent. A complete
linear stability theory has been carried out by \cite{PreziosiRosso} and \cite{Dijkstra92} who demonstrated that the background
shear can act to stabilize capillary or Rayleigh instability - this was subsequently seen experimentally by \cite{LowrySteen}. 
A weakly nonlinear theory that also incorporates the effect of insoluble surfactants on the interface has been carried out by
\cite{BBP2012} who derive a system of evolution equations reminiscent of the KS equation and extending the quiescent linear
stability study of 
\cite{Pozrikidis2001a}. 
The current study aims in extending the thin film study of \cite{Hammond1983,LRKCJ2006} into the fully nonlinear regime 
allowing the interface to approach either of the bounding concentric walls. 
Furthermore our problem setup is the limiting case of \cite{PreziosiRosso} when the background shear is vanishingly small and capillary forces
dominate (the present study is nonlinear however). In general our model provides a basis for further studies that can include 
surfactants or thermal effects as in~\cite{Pozrikidis2001a, CMP2002, CMReview}.
Finally we note that our study can provide insight into dominant two-phase capillary instabilities that have been used by
\cite{BP2008} to study particle encapsulation due to thread breakup - a string of particles along the thread axis
may behave as an inner rod in certain cases and thus affect the initial nonlinear dynamics according to flows analogous
to that studied here in the absence of electric fields.

Recently, \cite{Ding2013} investigated the linear stability of the electrified flow system when both inner and outer cylinders are present. As in \cite{Dijkstra92}, a constant pressure gradient is applied to drive the flow while the two immiscible fluids are treated as leaky dielectrics. 
The leaky dielectric model was first proposed in the work of \cite{Taylor66} to explain the steady deformation of a conducting drop, and demonstrated the importance of the electric tangential stresses. This was further developed by \cite{MelcherTaylor} and a comprehensive review of the model can be found in \cite{Saville1997}.  
Our interest lies in the description of nonlinear aspects of rod-annular flows and the linear assumptions of \cite{Ding2013} are dropped
while at the same time removing the background pressure-driven flow in order to make analytical progress with such dynamics.

Electrified core annular flows in the absence of a background base-flow have been studied recently by \cite{WP2011}, \cite{Wang2012} who find
that the electric field acts in a complex manner on the nonlinear interfacial dynamics.
For the interfacial electrohydrodynamics of a single fluid layer coating the inside of a vertical tube and driven by gravity \cite{WMP2012}, \cite{WPM2013a} studied the problem using a long wave theory and the leaky dielectric model. These nonlinear 
studies show that the electric field could either enhance or suppress the instability (see also the work by \cite{Ding2014} for a perfectly conducting layer which reached similar conclusions as in our previous studies in \cite{WMP09}). The setup of our current problem (see Figure \ref{domain}) is similar to the electrified film-coating problem mentioned above (\citet{WMP2012,WPM2013a}). The main difference is that we consider two 
leaky dielectric
viscous fluids confined in an annular region and as a consequence hydrodynamic and electrical fields must be determined
in both regions and coupled across the interface. In addition, the liquid layers can have thicknesses of the same order as the rod radius, so the thin film assumption is relaxed. A similar setup was considered by ~\cite{Reynolds1965} who investigated the 
linear electrohydrodynamic instabilities theoretically and carried out experiments. The experimental study used an AC field and interfacial instability was observed for low frequencies while stable patterns were observed at high frequencies. We note that the fluids used in the
Reynolds experiments were immiscible viscous silicone oil (Dow 200) and corn oil (Mazola), and such choices may raise a question regarding 
the allied linear stability work that neglected viscous effects. As pointed out by \citet{Taylor66} and  \cite{MelcherTaylor}, the tangential electric stress is vital and can only be balanced by viscous stresses at the interface. In addition, the analysis in~\cite{Reynolds1965} starts with the high frequency limit which leads to a system that is close to the perfect dielectric case. Eventually a Mathieu equation is derived to calculate the 
instability growth
rates of the system, and one correction term is included in the high frequency expansions to probe the effect of finite frequency. Similar analyses
can be found in \citep{GRC1999}, \cite{GGC2003} where AC fields are used with no electric tangential stress balances incorporated
into the models. Generally, then, the analysis of \cite{Reynolds1965} does not accurately bridge  
the theoretical and experimental predictions in the low frequency limit. In the current study we revisit the problem using a long wave 
nonlinear theory for the DC field  problem (equivalently the low frequency limit). We include viscous effects and 
subsequently show that our model is able to qualitatively predict the instability seen by Reynolds in his the experiments. 
Furthermore, we show that rich 
dynamics emerge by tuning certain electric field parameters such as the electric permittivity and conductivity, for instance. 
In addition to finite-time singular solutions in the form of interfacial spikes emerging away from the walls, 
axisymmetric patterning solutions with film thinning near the outer or inner tube wall are also discovered in certain parameter regimes, thus extending to three dimensions
the two-dimensional findings of \cite{CM05}. Interestingly, \cite{Reynolds1965} also discovered some (stationary) wavy pattern solutions in a high frequency electric field and with the interface remaining away from the tube walls. 
Indeed, previous studies on patterning under electric fields are overwhelmingly carried out in two dimensional geometries. 
We demonstrated here that even in the presence of the destabilizing capillary term in the interfacial curvature (this
is absent in two dimensions), patterning solutions still occur under fairly generic conditions.
A very different application of electrified two-phase rod-annular flows arises in the food processing industry that use such devices
to control processing with high-voltage pulsed electric fields - see for example \cite{Qin1994, Qin1995, JJH1999};
the electric field is believed to destroy microorganisms due to the induced electromechanical instability of cell membranes and is
considered to be a viable alternative to thermal treatments that can have global detrimental effects on treated substances.

The remainder of the paper is organized as follows. Section \ref{governingeqns} introduces the governing equations and boundary conditions
for the full axisymmetric problem. A long wave asymptotic approximation is carried out in section \ref{lweqns} and a system of nonlinear evolution equations is derived. The linear stability of the system is investigated in section \ref{linr} while nonlinear solutions are studied numerically and analytically
in section \ref{numerical}. Finally the results are summarized in section \ref{concl} where comparisons with experiments are also made.





\section{Governing equations}\label{governingeqns}
\begin{figure}
\centerline{ \includegraphics[width=4.in,height=2.in]{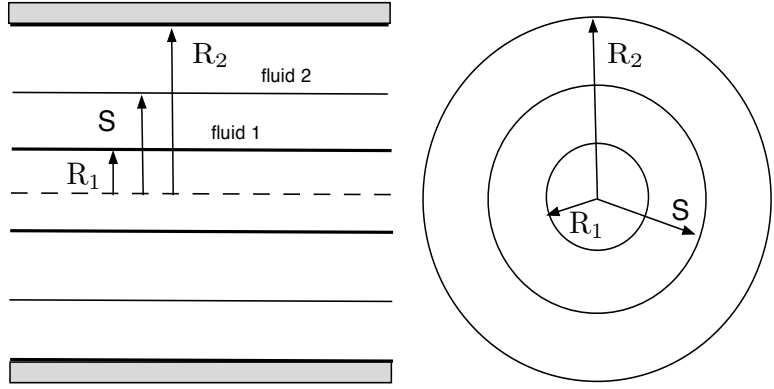} }
   \caption{Schematic of the problem. Inner and outer electrodes are placed at $r=R_1$ and $r=R_2$, between which are confined two immiscible fluids with densities $\rho_1,\rho_2$, viscosities $\mu_1,\mu_2$, electric permittivities $\epsilon_1,\epsilon_2$ and conductivities $\sigma_1,\sigma_2$.}
   \label{domain}
\end{figure}

We consider the axisymmetric evolution of a two-fluid rod annular flow confined between two cylindrical tubes as shown in Fig.~\ref{domain}. 
The inner and outer cylinders at $r=R_1$ and $R_2$ respectively,
serve as electrodes and a constant voltage potential difference is maintained between them by grounding the outer cylinder ($\phi_2=0$)
and keeping the inner one at $\phi_1=V_0$.
The equations and boundary conditions are non-dimensionalized by scaling 
lengths with the annulus thickness $d=R_2-R_1$, velocities by the capillary velocity $\gamma/\mu_2$, 
time by $d\mu_2/\gamma$, pressures by $\gamma/d$ and electric potentials by $V_0$. 
Here $\gamma$ is the coefficient of surface tension assumed to be constant.
For brevity we only write down the dimensionless equations.  

In the electrostatic limit appropriate to the present study, the electric fields are irrotational and given by $\mathbf{E}_i=-\nabla \phi_i$, where $i=1,2$ denotes regions $1$ and $2$, and $\phi_{i}$ is the voltage potential there. 
In addition, Gauss's law implies that ${\mathbf{\nabla}}\cdot(\epsilon_i\mathbf{E}_i)=0$ and so away from the interface the
potentials are governed by the axisymmetric Laplace equations,
\begin{align}
&\nabla^2\phi_{i}=0,\qquad\nabla^2\equiv\frac{\partial^2}{\partial r^2}+\frac{1}{r}\frac{\partial}{\partial r}+\frac{\partial^2}{\partial z^2}.
\end{align}\label{eqnlap}
The fluid dynamics are governed by the dimensionless Navier-Stokes equations
\begin{align}
\chi_iR_e \left(u_{it}+u_iu_{ir}+w_i u_{iz}\right)&=-p_{ir}+m_i\left(\nabla^2 u_i-\frac{u_i}{r^2}\right),\label{momer}\\
\chi_iR_e\left(w_{it}+u_iw_{ir}+w_i w_{iz}\right)&=-p_{iz}+m_i\nabla^2w_i,\label{momez}\\
 \frac{1}{r}\left(ru_i\right)_r+w_{iz}& =0,\label{cont}
\end{align}
where $R_e=\rho_2 \gamma d/\mu_2^2$, $\chi_2=1,\chi_1=\rho_1/\rho_2, m_2=1, m_1=m=\mu_1/\mu_2$ and subscripts denote partial derivatives.
The boundary conditions for the voltage potentials and no-slip at the walls $r=a=R_2/d$ and $r=b=R_1/d$, are
\begin{align}
& \phi_2(a,z)=0,\quad \phi_1(b,z)= 1\label{eq:phiwall}\\
&u_2(a,z,t)=w_2(a,z,t)=u_1(b,z,t)=w_1(b,z,t)=0.\label{eq:noslip}
\end{align}
At the interface $r=S$ we must satisfy
\begin{align}
&\phi_1(S,z)=\phi_2(S,z),\label{ebc1}\\
& q = Q\phi_{1n}(S,z)-\phi_{2n}(S,z),\label{ebcq}\\
&\phi_{1n}=R\phi_{2n},\label{ebc2}\\
&u_1(S,z,t)=u_2(S,z,t),\quad w_1(S,z,t)=w_2(S,z,t),\label{eq:contInterface}\\
& u_i(S,z,t)=S_t + w_i(S,z,t)S_z,\label{kin}
\end{align}
where $q$ is the surface charge density, $R, Q$ are the conductivity and permittivity ratios $R=\sigma_2/\sigma_1$ and $Q_1=Q=\epsilon_1/\epsilon_2$, $Q_2=1$ (the notations in \cite{Saville1997} are used). These boundary conditions represent
continuity of voltage potentials at the interface, Gauss's law, continuity of currents across the interface, continuity
of velocities, and kinematic conditions stating that the interface is a material surface.
Finally, the tangential and normal stress balances at interface are given by
\begin{align}
&\left[\frac{m_i}{1+S^2_z}\left(2S_z(u_{ir}-w_{iz})+(1-S_z^2)(u_{iz}+w_{ir})\right)\right]_1^2=\frac{E_b q(\phi_{1z}+S_z\phi_{1r})}{\sqrt{1+S^2_z}},\label{tsb}\\
&\left[-p_i + \frac{2m_i}{1+S^2_z}\left(S^2_zw_{iz}-S_z(u_{iz}+w_{ir})+u_{ir}\right) \right]_1^2=\nonumber\\
& \frac{1}{S\sqrt{1+S_z^2}}\left(1-\frac{SS_{zz}}{1+S_z^2}\right)-\frac{E_b}{1+S^2_z}\left[\frac{Q_i}{2}(1-S_z^2)(\phi^2_{ir}-\phi^2_{iz})-2Q_i S_z\phi_{iz}\phi_{ir}\right]_1^2,\label{nsb}
\end{align}
where $E_b=\epsilon_2 V_0^2/\gamma d$ and the jump notation $[(\cdot)]_1^2=(\cdot)_2-(\cdot)_1$ is used.
The model given above is the leaky dielectric model discussed originally by \citep{Taylor66, MelcherTaylor, Saville1997}.
It is interesting to note that on setting
$R=Q^{-1}$, equation (\ref{ebc2}) becomes $Q\phi_{1n}=\phi_{2n}$ which is the boundary condition across
an interface separating two perfect dielectric liquids - Gauss's law \eqref{ebcq} is dropped when dealing with perfect dielectrics.

\section{Long wave approximation}\label{lweqns}
\setcounter{equation}{0}

The problem posed above is highly nonlinear and in general requires direct numerical simulations.
However, the number of parameters is large and a full exploration of the dynamics using direct numerical
simulations is daunting. We make analytical progress by studying the nonlinear dynamics in the long wave limit
in order to derive more tractable reduced-dimension equations that can be analyzed and computed.
We assume that the axial length scale is much larger than the radial one and introduce a
slenderness parameter $\delta=\mathcal{R}/\mathcal{L} \ll 1$. 
Since there is no underlying flow (e.g. driven by gravity, an axial pressure gradient or the motion of one of the walls),
in the long wave limit an interfacial disturbance $r=S(z,t)$ will produce a flow that has the following
asymptotic expansions
\begin{align}
& u_i\sim \delta^2 u_i^0+\delta^4u_i^1+\cdots,\quad w_i\sim \delta w_i^0+\delta^3 w_i^1+\cdots\\
& p\sim p_i^0+\delta^2p_i^1+\cdots,\quad \phi_i\sim \phi_i^0+\delta^2\phi_i^1+\cdots.
\end{align}
In what follows we  only require solutions for the leading order
quantities hence we will drop the superscripts when writing expressions for such leading order solutions.

Beginning with the electrostatic problem and using the long wave scalings so that $\partial/\partial r =\mathcal{O}(1)$,
$\partial/\partial z =\mathcal{O}(\delta)$, the leading order Laplace equations become $\partial^2_r\phi_{1,2}^0+(1/r)\partial_r\phi_{1,2}^0=0$,
which can be readily solved using the boundary conditions \eqref{eq:phiwall}, \eqref{ebc1} and \eqref{ebc2} to find (superscripts are dropped)
\begin{align}
\phi_1&=\frac{R\ln(r/b)}{\ln(S/a)-R\ln(S/b)}+1,\label{phi1}\\
\phi_2&=\frac{\ln(r/a)}{\ln(S/a)-R\ln(S/b)}.\label{phi2}
\end{align}
As a check we note that if the inner layer is perfect conductor, i.e. $R\rightarrow 0$, the solution for $\phi_2$ recovers that found in the
study of \cite{WMP09}.

Next we consider the hydrodynamics and the coupling between the electric field and the fluid motion. 
As will be shown later, a slow timescale $t=\delta\tau$ is required in the kinematic condition in order
to retain unsteady effects at leading order. 
Therefore the inertia terms are small compared to other terms in \eqref{momer}-\eqref{momez}, as long as $\chi_i R_e$ is not too large. For example, the left hand side of (\ref{momez}) is of order $\chi_i R_e\delta^3$ while the right hand side has size $\mathcal{O}(\delta)$, which amounts to assuming $\chi_i R_e\ll \delta^{-2}$ in order to drop inertia. This ordering along with the scaling
$\partial_z\to\delta\partial_z$ enables us to make analytical progress and obtain closed form solutions at leading order - the
main balances are between the $r-$components of the viscous terms with the induced pressure gradients.
The axial leading order velocities $w_1$ and $w_2$ are obtained first (recall once more that superscripts are dropped), and the
corresponding radial velocities $u_1$ and $u_2$ follow from the scaled continuity equations \eqref{cont}.
The solutions satisfying no-slip at the walls are given by
\begin{align}
w_1 &= \frac{r^2-b^2}{4m}p_{1z}+ \frac{A_1}{m} \ln(r/b),\\
u_1 & = -\frac{(r^2-b^2)^2}{16mr}p_{1zz} - \frac{A_{1z}}{m}\left(\frac{r^2\ln(r/b)}{2r}-\frac{r^2-b^2}{4r}\right),\\
w_2 &= \frac{r^2-a^2}{4}p_{2z}+ A_2 \ln(r/a),\\
u_2 & = -\frac{(r^2-a^2)^2}{16 r}p_{2zz} - A_{2z}\left(\frac{r^2\ln(r/a)}{2r}-\frac{r^2-a^2}{4r}\right),
\end{align}
where the functions $p_i(z,t),\, A_i(z,t)$ are to be found. This can be done by using continuity of velocities
across the interface \eqref{eq:contInterface}, and the leading order contributions to the
tangential and normal stress balance equations  \eqref{tsb}-\eqref{nsb}. The leading order boundary conditions at $r=S$ become
\begin{align}
& w_1=w_2,\qquad u_1=u_2,\label{contu}\\
&w_{2r} -m w_{1r}= E_b\, q \,(\phi_{1z}+S_z\phi_{1r})|_{r=S_-},\label{tsb0}\\
& p_1-p_2 = \kappa - \frac{E_b}{2}\left(\phi^2_{2r}|_{r=S_+}-Q\phi^2_{1r}|_{r=S_-}\right),\label{nsb0}
\end{align}
where $\kappa=1/S-\delta^2 S_{zz}$ is the curvature retaining a high order regularizing term
(following \cite{EgDu1994,CM2006}). In order to ease the representation of \eqref{tsb0}, \eqref{nsb0} and subsequent
expressions affected by them, we introduce the following definitions
\begin{align}
E_T& = E_b\, q\, (\phi_{1z}+S_z\phi_{1r})|_{r=S_-},\qquad q=(Q\phi_{1r}|_{r=S_-}-\phi_{2r}|_{r=S_+}),\label{chargeq}\\
E_N& = \frac{E_b}{2}\left(\phi^2_{2r}|_{r=S_+}-Q\phi^2_{1r}|_{r=S_{-}}\right),\qquad \varkappa=\kappa-E_N.\label{eq:E_N}
\end{align}
In addition, all expressions above are expressed 
 in terms of the solutions (\ref{phi1}) and (\ref{phi2}).
Using the velocity continuity condition (\ref{contu}) and the leading order tangential stress balance (\ref{tsb0}), we obtain
\begin{align}
A_1& = -\frac{S^2}{2}\varkappa_z - S\,E_T + A_2,\label{a1}\\
A_2 &= \frac{(mF(S,a)-F(S,b))p_{2z}+4G(S,b)\varkappa_z+4S\ln(S/b)E_T}{4\ln(S/b)-4m\ln(S/a)},\label{a2}
\end{align}
where we have introduced the functions
\begin{align}
&F(S,\xi)= S^2-\xi^2,\\
& G(S,\xi) = \frac{1}{2}S^2\ln(S/\xi)-\frac{1}{4}(S^2-\xi^2),
\end{align}
and $\xi$ takes on the values $a$ or $b$.
The radial velocity continuity condition $u_1=u_2$ at the interface leads to the following relation between $p_2$ and $A_2$,
\begin{align}
&\frac{1}{16}\left[{mF^2(S,a)-F^2(S,b)}\right]\,p_{2zz} =\nonumber\\
 &\left[G(S,b)-mG(S,a)\right]A_{2z}+
\frac{1}{16}{F^2(S,b)}\,\varkappa_{zz}- G(S,b)\left(\frac{1}{2}{S^2\varkappa_z}+SE_T\right)_z.\label{pa2}
\end{align}
The evolution equation for $S$ can be obtained by substitution of $u_1, w_1$ (or $u_2$, $w_2$) 
into the kinematic boundary condition (\ref{kin}) and retaining leading order terms. As mentioned already a new slow timescale
$\tau=t/\delta$ (i.e. $\partial_t= \delta^2\partial_{\tau}$) needs to be introduced in order to have dynamics entering
at leading order and balancing with the nonlinear terms; the final equation becomes (written in conservative form)
\begin{align}
(S^2)_{\tau} + \left(\frac{F^2(S,a)}{8}p_{2z} + 2A_2G(S,a)\right)_z=0.\label{eqns}
\end{align}
Thus, the system that governs the evolution of
the interface $S(z,\tau)$ is given by (\ref{a2}), (\ref{pa2}) and (\ref{eqns}). An immediate consequence of \eqref{eqns}
is that it conserves the total mass of fluid - assuming periodic boundary conditions on an interval $L$ say, it follows
that $\partial_t\int_L S^2 (z,t) dz=0$ and so $\int_L S^2 dz={\rm constant}$.

When $m=0$, the system reduces to an equation describing the dynamics of a fluid layer attached to the interior of the outer cylinder,
\begin{equation}
8(S^2)_{\tau} + \frac{\partial}{\partial z}\left[-(S^2-a^2)^2\varkappa_{z} + 2(S^2\varkappa_z+2SE_T)\left(2S^2\ln(S/a)-S^2+a^2\right)\right]=0\label{eqnsm0}.
\end{equation}
Note that the limit $m\gg 1$ cannot be taken directly and the system needs to be rescaled properly to obtain an
equation analogous to \eqref{eqnsm0} but for a layer of fluid attached to the inner cylinder.
The limiting equation 
\eqref{eqnsm0} is the same as the one in \cite{CM2006} when the electric field is turned off; it also recovers the equation in \cite{WPM2013a} in the absence of gravity and the fast charge relaxation limit (namely, charge advection is neglected). 
Considering the case when the inner cylinder has a large radius, our equations are expected to recover the ones studied in \cite{CM05} in the large conductivity case. 
The following substitutions are then used (see also \cite{WPM2013a}),
\begin{align}
b\rightarrow \frac{B}{\epsilon},\quad a\rightarrow \frac{B}{\epsilon}+1, \quad S\rightarrow \frac{B}{\epsilon} + 
h,\quad r\rightarrow \frac{B}{\epsilon}+ y,
\end{align}
where $\epsilon$ is a small parameter and $B, h, y$ are order one quantities. Thus,
\begin{align}
q& = \frac{QR-1}{h-1-Rh} + O(\epsilon),\\
p_1-p_2 & =\varkappa = - \delta^2 h_{zz} - \frac{E_b}{2}\left(\frac{1-QR^2}{(1-h+Rh)^2}\right) + O(\epsilon),
\end{align}
which match equations (35) and (56) in \cite{CM05} despite a different scaling for the surface charge and film height; \cite{CM05} used the
free space electrical permittivity $\epsilon_0$ in their non-dimensionalizations and have a dimensionless channel width equal to $1+\beta$ (their notation), whereas we used $\epsilon_0\epsilon_2$ and a dimensionless gap width $d$. In addition, the variables $c_5, c_7$ and $p_2$ in their notation match $A_1, A_2$ and $p_2$ in our equations (\ref{a1}), (\ref{a2}), (\ref{pa2}). 
Due to the complex form of the general equation for arbitrary $m$, we only present the equation for the case $m=0$, $p_{1z}=0$. To leading order we find
\begin{align}
h_{\tau} + \left(\frac{1}{3}(h-1)^3p_{2z} + E_b (QR-1)R\frac{(h-1)^2h_z}{2(-h+1+Rh)^3}\right)_z= 0,
\end{align}
and this matches equation (43) in \cite{CM05} when the conductivities are large.


\section{Linear stability analysis}\label{linr}
\setcounter{equation}{0}
A perfectly cylindrical interface provides the following exact solution to the equations,
\begin{align}
&\overline{S} = c,\qquad \overline{p}_1-\overline{p}_2= \frac{1}{c} - \frac{E_b}{2c^2}\frac{\left(1-QR^2\right)}{(\ln(c/a)-R\ln(c/b))^2}, \qquad \overline{E}_T=0
\end{align}
where $c$ represents the unperturbed interface; recall that $b<c<a$, and using our non-dimensionalization we have $a-b=1$. 
We perturb  the base state by writing
\begin{equation}
S=c+S',\qquad p_2=\overline{p}_2+p',\qquad A_2=0+A',
\end{equation}
where $|S'|, |p'|, |A'|\ll 1$,
and use the linearized form of  (\ref{a2}), (\ref{pa2}) and (\ref{eqns}) to obtain
\begin{align}
&A'=\frac{(mF(c,a)-F(c,b))p'_z+4G(c,b)\varkappa'_z+4c\ln(c/b)E'_T}{4\ln(c/b)-4m\ln(c/a)},\label{la}\\
&p'_{zz}= \frac{16(G(c,b)-mG(c,a))A'_z + F^2(c,b)\varkappa'_{zz}- 8G(c,b)\left(c^2\varkappa'_{zz}+2cE'_{Tz}\right)}{mF^2(c,a)-F^2(c,b)},\label{lp}\\
&2cS'_t + \frac{F^2(c,a)}{8}p'_{zz} + 2G(c,a)A'_z=0,\label{ls}
\end{align}
where $\varkappa'_z = -S'_z/c^2 - \delta^2 S'_{zzz} - E'_{Nz}$ with
\begin{align}
E'_{Tz} &= \frac{E_bR (R Q-1)\ln(b/a)}{c^2\left(\ln(c/a)-R\ln(c/b)\right)^3}\,S'_{zz},\label{fetz}\\
E'_{Nz} & = -E_b(1-QR^2)\frac{\left(\ln(c/a)-R\ln(c/b)+1-R\right)}{c^3\left(\ln(c/a)-R\ln(c/b)\right)^3}S'_z.\label{fenz}
\end{align}
Normal modes of 
(\ref{la}), (\ref{lp}) and (\ref{ls}) are sought by writing
\begin{equation}
S'=\hat{S}\, e^{ikz+\lambda \tau},\quad p'= \hat{p} \,e^{ikz+\lambda \tau},\quad A'= \hat{A}\, e^{ikz+\lambda \tau}
\end{equation}
where hat quantities are constants and $\lambda$ is the growth rate - if $\mathcal{R}(\lambda)>0$ we have instability.
A straightforward calculation casts the eigenvalue problem for $\lambda(k)$ into $det(\bold{M})=0$ where $\bold{M}x=0$ with $x=[S',p',A']^T$;
the entries of the matrix 
$\bold{M}$ are given in Appendix \ref{lwM}. 
In the absence of the electric field the results agree with \cite{Pozrikidis2001a,CM2006} in the long wave regime. 
For completeness and for comparison purposes, the linear linear stability of the full axisymmetric problem is given in Appendix \ref{fullM}. 
In order to compare the results based on the long wave approximation and the full problem, the true wavenumber and growth rate from the former are scaled as $\delta k$ and $\delta^2\lambda$ respectively. Following \cite{CM2006}, the $\delta^2 S_{zz}$ term is retained and 
to fix things the numerical value $\delta=0.1$ is taken in current study. 
Other choices of small $\delta$ have also been tested and agreement is equally good (not shown for brevity). 

\begin{figure}
\centerline{ \includegraphics[width=5.in,height=2.6in]{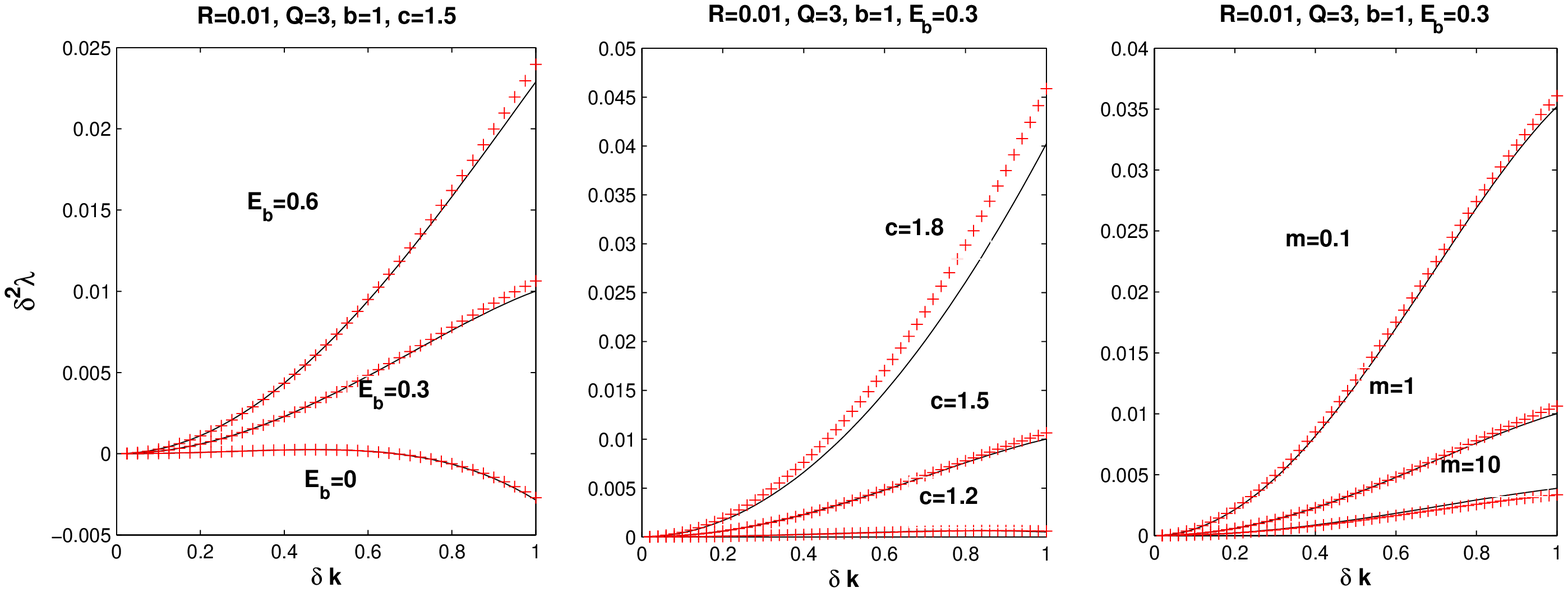} }
   \caption{Comparison of growth rates calculated from the long wave model (solid curves) and the full axisymmetric problem (crosses - red online). 
   Left panel - growth rates for fixed $c=1.5, m=1$ with varying $E_b$ as shown; middle panel - growth rates for 
   fixed $E_b=0.3$, $m=1$ with $c$ varying; right panel - growth rates for fixed $E_b=0.3$, $c=1.5$ with 
   $m$ varying. The remaining parameters are chosen to be $R=0.01, Q=3, b=1, a=2, \delta=0.1$.}
   \label{disp1}
\end{figure}

When an electric field is present the stability characteristics become much more complex, with the dimensionless
undisturbed interface radius $c$
and the electric field strength $E_b$ becoming important parameters.
We first consider the case of a perfectly conducting core, i.e. $R\rightarrow 0$, for which the tangential electric force $E'_{Tz}$ vanishes and 
\begin{align}
\varkappa'_{zz} \sim  -\frac{S_{zz}}{c^2}\left(1-E_b\frac{\ln(c/a)+1}{c\ln^3(c/a)}\right),
\end{align}
which indicates a critical interface position at $c=a/e\approx a/2.718$ around which the effect of the electric field can be reversed. This is consistent with the results in \citet{WMP09,WP2011} in their studies of perfectly conducting liquid jets. 
In the opposite limit of the annular fluid 2 being a perfect conductor, we have $R\rightarrow \infty$, and analogous results hold
\begin{align}
\varkappa'_{zz} \sim  -\frac{S_{zz}}{c^2}\left(1-E_bQ\frac{\ln(c/b)+1}{c\ln^3(c/b)}\right),
\end{align}
which gives a critical position $c=be=(a-1)e$.  For the general leaky dielectric case, we fix $m=1$ in order to reduce
the number of parameters yet provide a full picture of the complexity. 
The linearized equation (dropping the $\delta^2 S_{zz}$ term) becomes
\begin{align}
2cS'_{\tau}=F_0 \left(F_1 G_1 - F_2 G_2\right) S'_{zz},\label{eq:slin}
\end{align}
where 
\begin{align}
F_0 & = \left(\frac{F^2(c,a)}{8}+\frac{G(c,a)(b^2-a^2)}{2\ln(a/b)}\right)\frac{1}{\frac{b^2-a^2}{16}\left(-a^2-b^2+\frac{a^2-b^2}{\ln(a/b)}\right)},\\
F_1 & =\frac{F^2(c,b)}{16}+\frac{b^2-a^2}{4\ln(a/b)}G(c,b),\\
F_2 & = \frac{F(c,b)}{4}+\frac{b^2-a^2}{4\ln(a/b)}\ln(c/b),\\
G_1 & = \frac{1}{c^2}-E_b(1-QR^2)\frac{\ln(c/a)-R\ln(c/b)+1-R}{c^3(\ln(c/a)-R\ln(c/b))^3},\\
G_2 & = E_b\frac{R(RQ-1)\ln(b/a)}{c^2(\ln(c/a)-R\ln(c/b))^3}
\end{align}
Hence, stability depends on the sign of $F_0\left(F_1G_1-F_2G_2\right)$ above, with instability found
when this quantity is negative. The study of the effect of the relative position and electric strength can be found in \cite{WPM2013a} and 
in what follows we go on to explore numerically the effect of $R$ and $Q$ on the stability of the system. 
\begin{figure}
\centerline{ \includegraphics[width=6.in,height=3.7in]{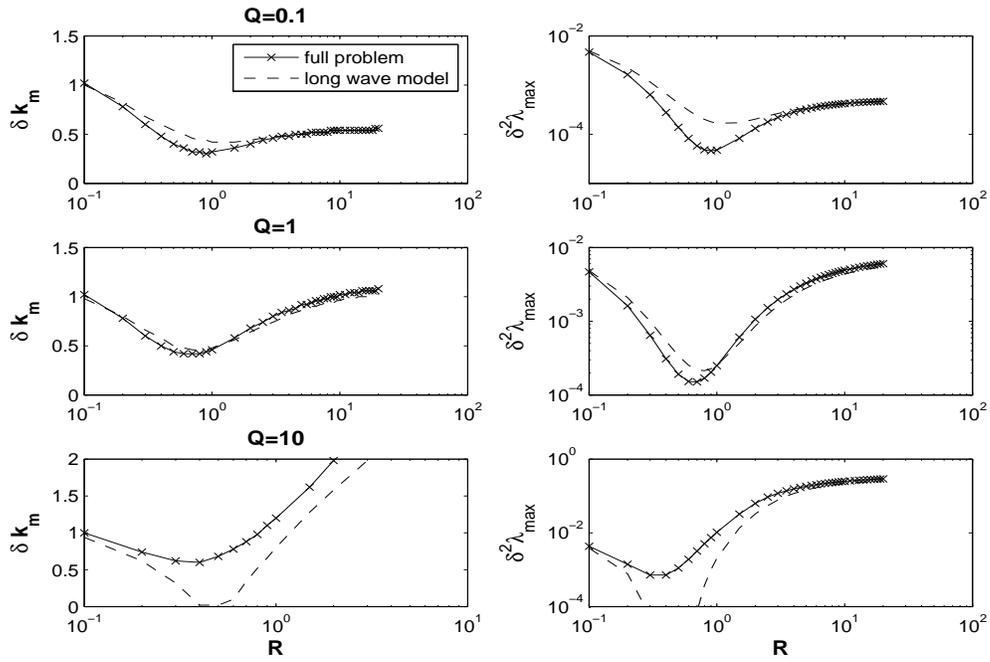}}
   \caption{The effect of $Q$ and $R$ on the maximum growth rates $\delta^2 \lambda_{max}$ and the associated wave numbers $\delta k_m$ for fixed parameters $b=1, a=2, \delta=0.1, m=1, E_b=0.3$. The solid curves are calculated from the full axisymmetric problem while the dashed ones are from our long wave model.}
   \label{disp2}
\end{figure}

Fig.~\ref{disp1} shows a comparison of stability characteristics between the long wave model \eqref{eq:slin} (solid curves)
and the full problem given in Appendix \ref{fullM} (crosses). As mentioned earlier, in order to compare the two we need to plot long wave results using unscaled variables - in this
particular set of results the wavenumber  $\delta k$ and the  growth rate 
$\delta^2 \lambda$. There are eight dimensionless parameters in the problem (including $\delta$) and we provide results
that fix the conductivity ratio to $R=0.01$, permittivity ratio to $Q=3$, dimensionless inner electrode radius to $b=1$,
dimensionless outer electrode radius to $a=2$ and $\delta=0.1$. The remaining parameters are the dimensionless electric field
strength $E_b$, the undisturbed jet radius $c$ and the inner to outer fluid viscosity ratio $m$. The left panel of Fig.~\ref{disp1}
in turn fixes $c=1.5$ and $m=1$ and varies $E_b$; the middle panel fixes 
$E_b=0.3$, $m=1$ and varies $c$, while the right
panel fixes $E_b=0.3$, $c=1.5$ and varies $m$. Firstly, the results show that there is excellent agreement between the
long wave model and the full linear theory for small wave numbers, and it is important to also note that the agreement
persists for moderate order one values of $\delta k$ also.
For the small $R=0.01$ used and $Q=3$, it is seen from the left panel that as the electric field increases the band
of unstable modes also increases.  The middle panel considers the effect of the undisturbed interface position, all other parameters fixed
as above but with $E_b=0.3$; it is seen that as the outer layer becomes thin (this happens as $c\to 2$, hence the curve having $c=1.8$
in the figure) the instability is enhanced. This can be understood physically since $R=0.01$ and so the outer thin layer is also
poorly conducting relative to the inner fluid and so a large electric field is set up across it that enhances the instability.
Finally, the right panel considers the effect of the inner to outer fluid viscosity ratio $m$ for fixed $c=1.5$ and $E_b=0.3$; 
a more viscous outer layer is seen to give larger growth rate than the case with comparable viscous layers or a more viscous inner layer.

Next we consider the effect of the electrical properties of the fluids $R$ and $Q$ on the stability - see Fig.~\ref{disp2}.
To do this we fix $b=1$, $a=2$, $m=1$, $\delta=0.1$ and $E_b=0.3$ and calculate the variation of the maximally
unstable wavenumber $\delta k_m$ (left panels) and the corresponding maximum growth rate $\delta^2 \lambda_{max}$ (right panels)
with $R$ and for three different
values of the permittivity ratio $Q=0.1,\,1,\,10$.
The dashed lines are from the long wave model while the solid-cross lines are from the full axisymmetric problem of Appendix \ref{fullM}. 
The first two rows of Fig.~\ref{disp2} for $Q=0.1$ and $Q=1$ respectively, show that the long wave model captures both $\delta k_m$ and $\delta^2\lambda_{max}$ quite well compared to the full problem, even for moderate wave numbers. 
Note however, that the long wave model overestimates the growth rate a little when $R$ is close to one. 
For the last row of Fig.~\ref{disp2} having $Q=10$, the trend for both $\delta^2\lambda_{max}$ and $\delta k_m$ is captured qualitatively.
The long wave model appears to support a small window of stable values of $R$ close to one, and this is not seen in the full problem. 
Proceeding to higher order terms in the long wave model may resolve the issue - we only kept $\delta^2 S_{zz}$ terms in the 
interface curvature but not in other terms -
 (see the discussion in \cite{Calvo2005} also). 
Our results show, however, that for most cases having moderate $Q$ and either large or small $R$, the agreement of the long wave prediction is reasonably good. 

In the remainder of this study we focus on the nonlinear stages of the
instability by studying in detail the dynamics of the long wave model and ensuring that we are consistent with the linear findings
and the limitations of the model when $Q$ is large.


\section{Numerical simulation of the nonlinear evolutions}\label{numerical}
\setcounter{equation}{0}
Our main interest is in the nonlinear electrohydrodynamic effects and due to the large number of parameters present in
the derived mathematical models we will concentrate on cases when $m=1$, i.e. the two fluid viscosities are the same.
The effects of viscosity contrast on thread or layer stabilities can be found in \citep{StoneLeal89a, Pozrikidis99, CMP2005, BPSW2013,Wang2013} for example, where they are observed to affect growth rate values rather than the qualitative features of nonlinear phenomena.
We anticipate, therefore, that results for $m=1$ will be similar to the cases with $m\sim O(1)$. We also note that the
study of the extreme case $m\rightarrow \infty$, i.e. a single electrified layer with DC field, can be found in \cite{WPM2013a}. 

\subsection{Evolution equation for $m=1$}

Substitution of $m=1$ into (\ref{a2}), (\ref{pa2}) and (\ref{eqns}) yields the following equation for the interface position $S(z,\tau)$,
\begin{align}
&(S^2)_{\tau}+\left[\left(\frac{F^2(S,a)}{8}+\frac{G(S,a)(b^2-a^2)}{2\ln(a/b)}\right)p_{2z}+\frac{2G(S,a)}{\ln(a/b)}\left(G(S,b)\varkappa_z+S\ln(S/b)E_T\right)
\right]_z=0\label{sm1}
\end{align}
where $\varkappa=\kappa-E_N$ and
\begin{align}
p_{2z}&=C\left[\left(\frac{b^2-a^2}{4\ln(a/b)}G(S,b)+\frac{F^2(S,b)}{16}\right)\varkappa_z +\left(\frac{(b^2-a^2)\ln(S/b)}{4\ln(a/b)}+\frac{F(S,b)}{4}\right)SE_T\right] + f(t)\label{pm1},\\
C & = \frac{16}{b^2-a^2}\left(-a^2-b^2+\frac{a^2-b^2}{\ln(a/b)}\right)^{-1},\\
E_T &= E_b(QR-1)R\frac{\ln(b/a)}{S^2\left(\ln(S/a)-R\ln(S/b)\right)^3}S_z,\label{fet}\\
E_N &= \frac{E_b(1-QR^2)}{2} \frac{1}{S^2\left(\ln(S/a)-R\ln(S/b)\right)^2}.\label{fen}
\end{align}
In (\ref{pm1}) the term $f(t)$ represents a quasi one-dimensional force due to an integration in space 
(see also \citep{Renardy94,Renardy95, DTP1995b, DTP1995a} for some discussion in related problems). 
In what follows this is fixed by the initial conditions to be zero.


To carry out the transient and large-time simulations we used the PDE solver EPDCOL (see \cite{Keast1991}) which utilises finite element 
discretizations in space and advances the system in time using Gear's method. 
These algorithms have been used successfully in related studies in solving liquid thread problems \citep{CMP2002, Conroy2011PF,Wang2012,Wang2013}.
The accuracy of the code was tested by comparing the numerical results with small initial conditions with the analytical
ones provided by linear theory. For
the results reported here we take as initial conditions
\begin{equation}
S(z,\tau=0)= b +(c-b)(1+0.05\cos(2\pi z/L)),
\end{equation}
along with the following with the following boundary conditions that ensure periodicity
\begin{equation}
S_z(-L/2,t)=S_z(-L/2,t)=0,\quad S_{zzz}(L/2,t)=S_{zzz}(L/2,t)=0.
\end{equation}

\subsection{The case of no electric field, $E_b=0$}\label{nofield}
\begin{figure}
  \centerline{ \includegraphics[width=6.in,height=5.2in]{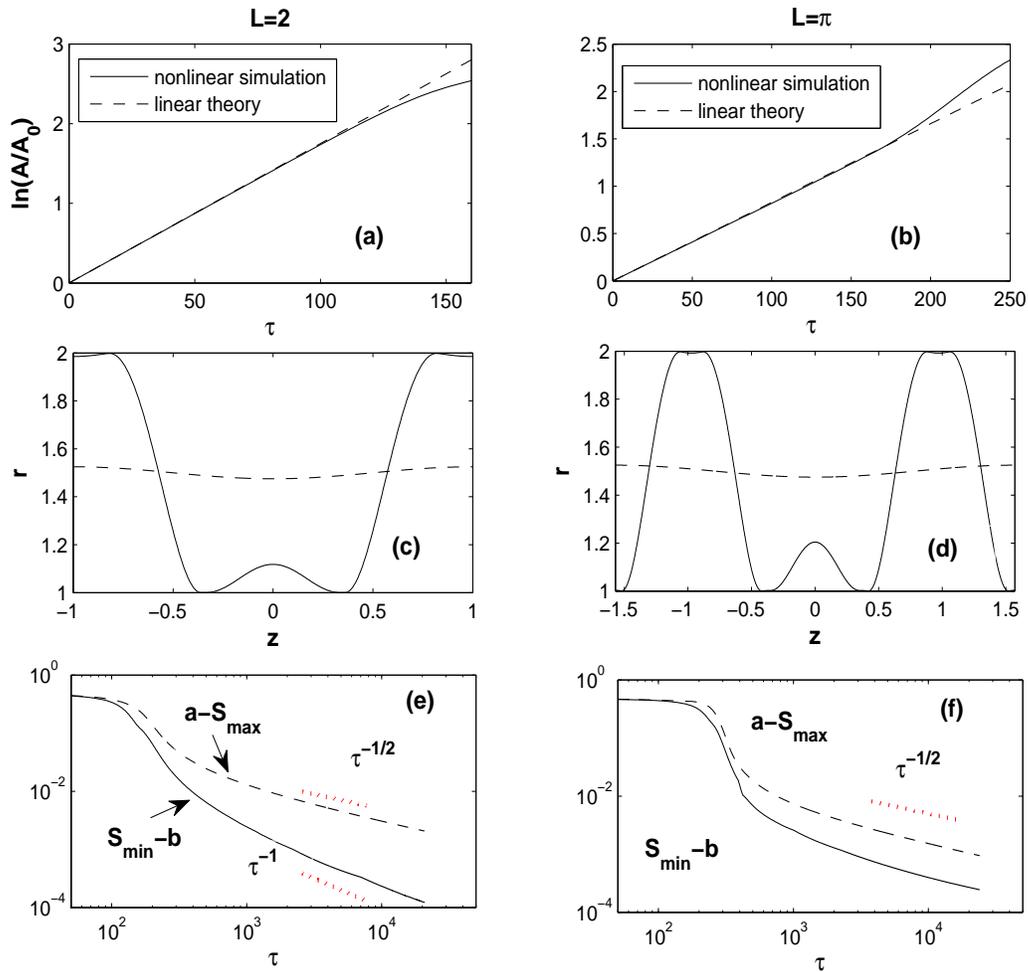} }
   \caption{Simulation results for $a=2,b=1,c=1.5, \delta=0.1$ without electric field, $E_b=0$. The left column shows results for $L=2$ or $k=\pi$ while the right column corresponds to $L=\pi$ or $k=2$.
   Panels (a), (b) show the comparison between the linear theory and nonlinear simulations. Panels (c), (d) show the initial interface position with
    dashed curves and the position at $\tau=10^4$ with solid curves. The upper ($r=2$) and bottom ($r=1$) box boundaries 
    represent the position for the outer and inner cylindrical walls. The evolution of $(S_{min}-b)$ and $(a-S_{max})$ is shown in panel (e) and (f) as indicated in figure.}
   \label{eb0}
\end{figure}

In this case our results also recover the film drainage solutions found in the literature - \citet{Hammond1983,LRKCJ2006}. 
We find that the fluid interface tends to asymptotically touch the tube wall 
and the minimum film thickness scales as $\tau^{-1}$ for a collar next to another collar, or 
$\tau^{-1/2}$ for a lobe next to a collar, using the nomenclature of \cite{LRKCJ2006}. 
This is to be expected when one layer is asymptotically thin so that the Hammond equation holds. 
To confirm this limit we write
$S=b+\eta$ where $|\eta|\le b$, and substitute into equation (\ref{sm1}) to find the following equation, to leading order,
\begin{align}
\eta_{\tau} + \frac{1}{3}\left(\eta^3\left(\frac{\eta_z}{b^2}+\delta^2\eta_{zzz}\right)\right)_z=0.\label{hameqn}
\end{align}
A similar equation with $b$ replaced by $a$ is derived when the upper layer is thin instead.
These results are consistent with \cite{Hammond1983}, \cite{CM2006} since the leading order equation neglects the 
curvature of the cylinder surfaces and is valid for a fluid layer coating either the interior or exterior surface.

In Fig.~\ref{eb0} we present results of numerical computations of \eqref{sm1} with $E_b=0$ and for 
$c=1.5, a=2, b=1$, i.e. the outer and inner cylinder radii are $a=2$ and $b=1$, respectively, while the unperturbed cylindrical interface is
half way in between and has
radius $c=1.5$. Two different domains are considered, $L=2$ and $L=\pi$ (equivalently different perturbation
wave numbers $k=\pi$ and $k=2$), and results are given to describe the evolution of the interface in each case.
In the early stages of the evolution depicted in panels \ref{eb0}(a) and \ref{eb0}(b), we see
excellent agreement between linear theory (dashed) and the simulation results (solid curves), for $L=1$ and $L=\pi$ as shown. 
The deviation occurs at $\tau\approx 120$ and $\tau\approx190$, respectively, and these are the times
where nonlinear effects enter. 
(Note that to ensure the code's accuracy, the mass $\int_{-L/2}^{L/2}S^2dz$ is monitored and found to be
conserved with an error of typically at most $10^{-4}$ - the same holds for the cases with the electric fields in the next sections.)
The long time evolution is given in panels (c) and (d) with the dashed curves representing the initial conditions and the solid curves the solution
at the end of the computation where $\tau>2\times 10^4$. The results show that the interface grows and reaches the outer and inner walls
simultaneously. We observe the
appearance of collar and lobe arrangements analogous to the findings of \citet{LRKCJ2006} for the Hammond equation
(there is only one bounding solid surface in that case); the formation of these is also found to be sensitive to the domain length and this is
also supported by our computations of the fully nonlinear systems. 
A more detailed understanding of the large time draining dynamics can be found by computing the minimum film thickness on
the upper and lower walls, respectively. Defining $S_{max}$ and $S_{min}$ to be the maximum and minimum interfacial positions provides
the minimum upper and lower film thicknesses to be $a-S_{max}$ and $S_{min}-b$, respectively. The evolution of these quantities
is plotted using logarithmic scales in panel (e) for $L=1$ and panel (f) for $L=\pi$. For the shorter domain our numerical findings
indicate that for large $\tau$ we have $S_{min}-b\sim \tau^{-1}$ while $a-S_{max}\sim \tau^{-1/2}$ - see superimposed dotted line
fits in Fig.~\ref{eb0}(e). This in turn implies that the adjacent collars at the lower cylindrical surface do not drain into each other, while
at the outer surface the lobe in the approximate region $-1.2< z< -0.8$ (or $0.8< z< 1.2$ due to spatial periodicity) 
drains into the collar approximately spanning
the region $-0.8< z< 0.8$; these findings are in lines with those reported in \citet{LRKCJ2006}.
When $L=\pi$ the domain is now sufficiently large to support additional small structures as seen in Fig.~\ref{eb0}(d).
In particular small lobes appear at both the outer and inner surfaces (in the vicinity of $z\approx \pm 1$, $z\approx \pm 0.4$ and $z\approx \pm 1.5$)
and these drain into the larger adjacent collars. Hence both outer and inner minimum film thicknesses are expected to scale as $\tau^{-1/2}$
and this is indeed confirmed by the numerical results in Fig.~\ref{eb0}(f).
We shall call such asymptotically singular solutions as touching solutions in the present article.
For longer time and longer domains, the detailed computations of \citet{LRKCJ2006} show the occurrence of more subtle dynamics,
such as the sliding of collars over small lobes - we have not attempted such computations and note that they are probably
beyond the capabilities of our current codes (times as large as $10^{12}$ need to be accessed).
The non-electrified results capture the phenomena described by \citet{LRKCJ2006} in the vicinity of wall touchdown after a fully
nonlinear evolution brings the interface simultaneously close to the inner and outer bounding surfaces. In what follows we
consider the electrified case and investigate in detail how the solution behaviour changes in response to the electric field.

The results in Fig.~\ref{eb0} were calculated by fixing the undisturbed interface position to be at $r=c=1.5$.
We have carried out a large number of computations to ascertain the behaviour of the flow when $c$ is either closer
to the outer wall at $a=2$ or the inner wall at $b=1$. Our simulations indicate strongly (results not shown) that if $c$ is
sufficiently close to one of the walls then a Hammond type solution develops with thinning near the closest wall (i.e.
one-sided touching in the nomenclature used above).
For the particular case of $a=2$ and $b=1$, we find that outer wall touching develops when $c\gtrsim 1.8$ while thinning at the inner wall takes place when $c\lesssim 1.3$.

\subsection{Electric field effects for perfect dielectric liquids}\label{PDcase}

For $E_b\ne 0$ the behaviour of solutions depends on  the electric parameters and numerical computations
are necessary. 
The simulations are stopped when the film thickness is typically less than $10^{-3}$ or $10^{-4}$. We first consider the perfect dielectric case
which follows by taking $R=Q^{-1}$ in equations \eqref{sm1}-\eqref{fen} - physically the electric tangential forces are 
absent and only the normal stress balance is modified by the Maxwell stresses.
Our simulations explored a large parameter space and we find that 
it is possible to encounter solutions that touch one of the walls in finite time and have cusp structures locally  
(as in \citet{WP2011}), as well as solutions that reach the walls asymptotically in time. 
A few representative results follow for a domain of length $L=2$  (qualitatively similar results are obtained for longer domains). 
\begin{figure}
  \centerline{ \includegraphics[width=5.in,height=3.8in]{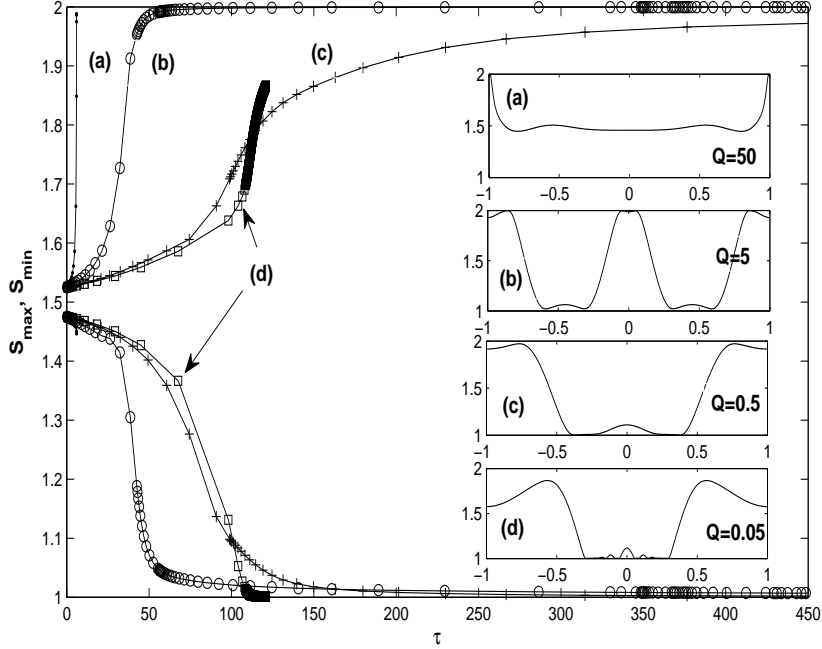} }
   \caption{Evolution of $S_{min}$ and $S_{max}$ for different values of $Q$ as indicated in the figure. 
   The insets show the fluid interface position at the end of the simulations, with $c=1.5, a=2, b=1$ and $E_b=0.3$ fixed. 
   Panel (a) represents a finite-time touching solution at the outer wall with a local cusp geometry. The other panels represent
   solutions that form asymptotically thinning films at the outer or inner walls as shown.}
   \label{pd1}
\end{figure}
\begin{figure}
  \centerline{ \includegraphics[width=5.in,height=3.8in]{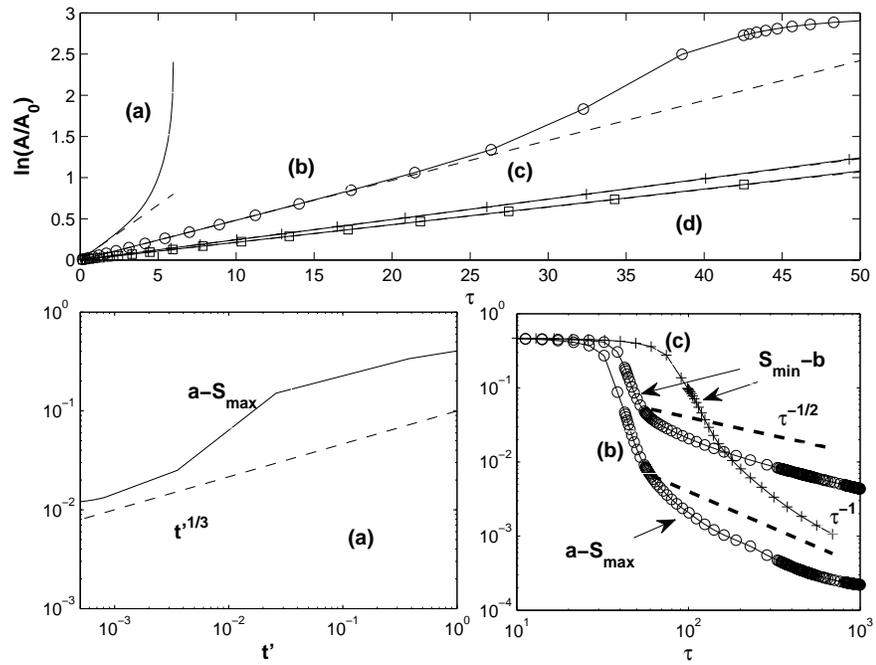} }
   \caption{Detailed evolution of results in Fig.~\ref{pd1}. Top panel - evolution of scaled amplitude along with comparison with linear theory
   at small times (dashed lines). Bottom left panel - scalings for finite-time singular solution case (a) confirming the $(t^\prime)^{1/3}$
   analytical prediction. Bottom right - numerical prediction of the analytical
   scalings for the film drainage confirming that $(S_{min}-b)\sim \tau^{-1/2}$, $(S_{min}-b)\sim \tau^{-1}$ in cases (b) and (c), 
   and $(a-S_{max})\sim \tau^{-1}$ for case (b).}
   \label{pd2}
\end{figure}

Numerical results are given in Figures~\ref{pd1} and \ref{pd2} for different values of the permittivity ratio $Q=50,\,5,\,0.5,\,0.05$. 
Recall that $Q=\epsilon_1/\epsilon_2$
is the ratio of inner to outer values, and for water/olive oil systems, for example, we obtain $Q\approx 25$ when the water wets the inner
electrode and $Q\approx 0.04$ when it wets the outer electrode. Fig.~\ref{pd1} shows the evolution of $S_{min}$ and $S_{max}$, the minimum
and maximum interface positions respectively, for different $Q$, and the insets depict the interfacial position at the last computed time.
The top panel of Fig.~\ref{pd2} shows the evolution of the scaled interfacial amplitude and superimposes the results of
linear theory (dashed lines). It can be seen that for larger $Q$ nonlinearity enters much sooner while for smaller values, e.g. $Q=0.5, \,0.05$
the linear results hold very accurately for the whole time interval depicted. 

The results suggest that the dynamics and ultimate
nonlinear structures that form at large times, are strongly dependent on $Q$. For large $Q$ (e.g. $Q=50$ in Fig.~\ref{pd1}) the interface
is driven to the outer electrode relatively quickly while at the same time staying away from the inner electrode. As can be seen from the
evolution of $S_{max}$ in the curve labelled (a), the interface touches the wall at a finite time $\tau\approx 5.97$,
and appears to be doing so with a sharp
cusp-like local geometry as seen from the last computed solution in the inset labelled (a) (such phenomena were also observed
by \citet{WP2011} in their related single fluid study, analogous to the annular region in the present problem being hydrodynamically
passive). The physical reason for this large $Q$ behaviour is that the electrostatic forces dominate over the capillary ones
and so outer wall touching is encountered before capillary instability can take over and push the interface towards
the inner wall electrode. As $Q$ decreases and the capillary forces become more prominent, the interface touches the inner and
outer electrodes simultaneously at first (see the results in Fig.~\ref{pd1} for case (b) having $Q=5$). For a lower value $Q=0.5$
the numerical results predict that the inner wall touchdown is more attractive as can be seen from the evolution of $S_{max}$ and $S_{min}$
in Fig.~\ref{pd1} and the associated final computed profile in the panel (c) inset. At lower values still, e.g. $Q=0.05$, our computations
had to be stopped long before the interface comes close to the outer electrode, at which time however it is almost touching the inner
wall and indeed some intricate collar lobe structures form on smaller length scales (see the insets (d) in Fig.~\ref{pd1}).

Details of the dynamics in the vicinity of touchdown events are considered in detail in the two lower panels of
Fig.~\ref{pd2} for $Q=50,\,5,\,0.5$ (labelled as cases (a), (b) and (c), respectively). The single outer wall touching event when $Q=50$
is considered in the left lower panel, and the results indicate that touching happens after a finite time $\tau_s$, say (as mentioned above
$\tau_s\approx 5.97$), with the minimum outer fluid layer thickness scaling as $a-S_{max}\sim (t^{\prime})^{1/3}$ where
$t^\prime=\tau_s-\tau$; the exponent $1/3$ is predicted from the logarithmic plot of the data as indicated on the figure - analytical
evidence of this is provided below.
Note that this in turn means that
the speed of approach diverges as $(t^{\prime})^{-2/3}$ as $t^\prime\to 0+$. These scalings confirm analysis of the equations near
the singular event as discussed by \cite{WP2011} in a simpler problem. Considering $Q\gg 1$ in equations \eqref{sm1}-\eqref{fen}, we see
from \eqref{fen} that in this limit we have $E_N\sim E_b/(2S^2\ln^2(S/a))$, and  
setting $\xi(z,\tau)=a-S(z,\tau)$ with $0<\xi\ll 1$ in \eqref{sm1}, yields to leading order
\begin{equation}
\xi_{\tau} + \frac{1}{3}\left(\xi^3\left(\frac{\xi_z}{a^2}+\delta^2\xi_{zzz}\right)\right)_z + \frac{E_b}{3}\xi_{zz}=0.\label{spikefilmeqn}
\end{equation}
In addition to $\xi\ll 1$ we also have $t^\prime=(\tau_s-\tau)\ll 1$ and in addition the solution focusses around a point $z=z_s$.
Hence the derivatives $\partial_\tau\sim (t^{\prime})^{-1}$ and $\partial_z\sim (z-z_s)^{-1}$ are asymptotically large as touchdown takes place
and so the following balances must hold
\begin{equation}
\xi_{\tau}\,\sim\,(\xi^3\xi_{zzz})_z\,\sim\,\xi_{zz}.
\end{equation}
This enables the determination of the scalings
from which a set of scalings can be derived
\begin{equation}
\xi\sim (\tau_s-\tau)^{1/3},\quad z-z_s\sim (\tau_s-\tau)^{1/2}.
\end{equation}
The numerical results presented in 
Fig.~\ref{pd2} are consistent with the analytical findings.

Details of the touchdown dynamics for $Q=5$ and $0.5$ are provided in Fig.~\ref{pd2}, bottom right panel, where we now
emphasize that the interface reaches the walls asymptotically in time and so the limit $\tau\gg 1$ is the appropriate one.
For $Q=5$, labelled case (b),  we tracked both $a-S_{max}$ and $S_{min}-b$ (open circles are used for $Q=5$)
as they go to zero asymptotically, and the logarithmic plots
indicate that the outer layer dynamics are faster and scale as $\tau^{-1}$ whereas the inner layer thins according to the
slower rate $\tau^{-1/2}$. These findings are analogous to the simultaneous asymptotic touchdown found in the absence
of electric fields - see Fig.~\ref{eb0}. At a smaller value of the outer to inner permittivity ratio $Q=0.5$, we find one-sided
touching at the inner wall with a thinning rate given by $S_{min}-a\sim\tau^{-1}$ as seen from the results indicated by crosses.
Once again, the capillary forces dominate over the electrostatic ones to produce one-sided touching analogous to Hammond dynamics.
For the smallest value considered $Q=0.05$, 
our code was not able to track the evolution to very long times due the emergence of multiscale
solutions due to the presence of small annular rings forming along the inner wall. 
The evolution of $S_{min}$ in Fig.~\ref{pd1} strongly suggests that we have an inner wall one-sided touching solution but our data
is not reliable enough to predict a $\tau^{-1}$ thinning rate. Analytically, however, we observe that as
$Q\rightarrow 0$ in expression (\ref{fen}) we obtain, to leading order, $E_N\sim -\frac{E_b}{2}\frac{Q}{S^2\ln^2(S/b)}\ll 1$ and so the
normal contribution to the Maxwell stresses vanishes (recall that we are considering $R=0$ in this section and hence the tangential
contribution $E_T$ is also absent). This observation analytically confirms our physical interpretation of the one- versus two-sided
touching with the former dominating when capillary forces are larger than electrical ones.

\subsection{Electric field effects for leaky dielectric liquids}\label{LDcase}

\begin{figure}
  \centerline{ \includegraphics[width=6.in,height=4.in]{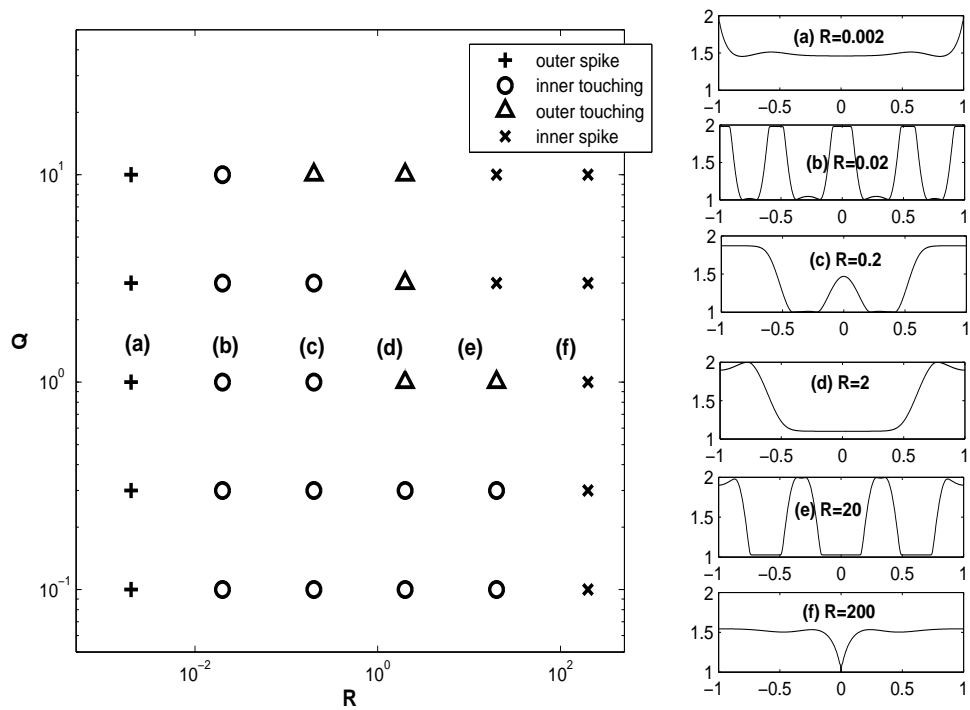} }
   \caption{Phase diagram depicting the different breakup behaviour in $Q$ - $R$ space when $c=1.5, a=2, b=1, E_b=0.3$ are fixed. The 
   symbols $+$ ($\times$) represent the finite time singular solution that touches the outer (inner) tube wall; the symbols
   $\bigcirc$ ($\triangle$) represent the film draining solutions occurring at the inner (outer) tube wall. 
   The insets show the interface position towards the end of the numerical simulation for individual cases when $Q=1$ and various $R$
   shown on the figure. In contrast to the non-electrified or the perfect dielectric case, no touching solutions that occurs simultaneously at 
   the inner and outer walls have been found.}
   \label{dce}
\end{figure}

For the general case of leaky dielectric liquid layers, the normal and tangential stresses $E_N$ and $E_T$ in system
\eqref{sm1}-\eqref{fen} are non-zero and enter to influence the dynamics. The equation was solved numerically for arbitrary
values of the permittivity and conductivity ratios $Q$ and $R$, and we
begin by showing representative results for a domain of length $L=2$, geometrical parameters $a=2$, $b=1$, $c=1.5$ as in previous sections,
and a value $E_b=0.3$ for the electric parameter measuring the strength of the applied electric field.
Results from $35$ computations that follow solutions to either finite-time or asymptotic touching are collected in the phase diagram Fig.~\ref{dce}, that
classifies solutions in $R$-$Q$ space according to their terminal or large time characteristics. 
In particular different symbols are used to denote a spike solution touching the outer wall in finite-time ($+$ symbols), inner wall touching
at large times ($\bigcirc$ symbols), outer wall touching at large times ($\triangle$ symbols), and inner wall touching spike solution in finite time
($\times$ symbols).
The values of $R$ and $Q$ investigated span several orders of magnitude in order to make our results
representative of realistic physical systems.
We note that for longer domains similar results are found
with the same trends of finite time touching or asymptotically thinning solutions.

Several noteworthy features emerge from these results. First, when $R$ is small or large, corresponding to a highly conductive
inner or outer layer, respectively, we find that a spike finite-time singularity solution forms relatively quickly and touches the outer
or inner wall depending on the value of $R$. Such results can be seen in the first column of Fig.~\ref{dce} that has $R=0.002$, with solutions
evolving to outer wall finite-time touchdown, and also analogous results in the last column of the figure for $R=200$ 
but with inner wall finite-time touching in this case.
We emphasize that for small or large $R$, the solution features and general structure of terminal states are unaffected by the value of $Q$.
Typical terminal solutions are included in the panel insets (a) and (f) that clearly show the outer and inner wall cusp-geometry touching
solutions, respectively. Physically, these terminal states tend to separate regions of more/less conductive fluids (depending
on whether for $R$ is small/large) into annular rings that are attached to one of the walls.
These breakup singular solutions are consistent with the experimental results of \cite{Reynolds1965} who
reports spike solutions emerging for a very low frequency AC field with two very viscous leaky dielectric oils 
arranged between two concentric cylinders. According to \cite{Reynolds1965}, the experiments were carried out 
for $Q\approx 1.24$ and $R\approx 0.01$ using our notation, and these values are fully consistent with our phase diagram Fig.~\ref{dce} in that the outer spike solution is expected for small $R$ as shown. 
Other parameters used in the experiments are that
the surface tension $\gamma\approx 1\ {\rm dyn/cm}$ and the annulus thickness $d\approx 3\ {\rm cm}$.
The experiments used Dow 200 silicone oil  and Mazola corn oil for the two phases, and so we can estimate values of the density
$\rho\approx 1\ {\rm g/cm^3}$ and viscosity $\mu\approx 1\textendash 500\ {\rm g/(cm^2s)}$ to provide an estimate
for the Reynolds number
$R_e\approx 1\textendash 10^{-4}$; such values are consistent with our assumption $\chi_i R_e\ll \delta^{-2}$ for order one $\chi_i$. 
The qualitative agreement of our results with the experiments is encouraging and provides physical justification for our models.

For moderate values of the conductivity ratio in the range $0.02\le R\le 20$, we find that one-sided wall touching prevails.
This is in contrast to the non-electrified and perfect dielectric computations for the same geometrical parameters - see
Fig.~\ref{eb0} and \ref{pd1} where simultaneous inner and outer wall touching was found, albeit
at different asymptotic rates.
More specifically, for this range of $R$ and small to moderate values of $Q$, we see that inner wall touching is most prominent.
This happens for all investigated values of $Q$ for $R=0.02$, whereas for $R=0.2$ inner wall touching gives way to outer wall touching 
for $Q\ge 10$. As $R$ increases a switch to outer wall touching occurs for smaller $Q$ - e.g. for $R=2$ outer wall touching happens
for $1\le Q\le 10$. At the larger value of $R=20$ we transition from outer wall touching at $Q=1$ to an
inner wall touching spike solution in finite-time for $Q\ge 2$.

These results can be understood in physical terms by considering the relative importance of electrostatic and capillary forces.
Recall that $R=\sigma_2/\sigma_1$ represents the conductivity ratio of outer to inner fluids, and so $R\ll1 $ implies that the
inner layer is perfectly conducting whereas $R\gg 1$ corresponds to the outer layer being perfectly conducting, to leading order.
In the former case the field acts in the outer region alone and the Maxwell stresses are sufficiently strong to overcome
capillary instability (that tends to produce inner wall touching) and induce outer wall cusp-like touching as seen in the results
for $R=0.002$. If $R\gg 1$, however, the field mostly acts in the inner region and by similar reasoning the Maxwell stresses
there work in tandem with the capillary instability to produce cusp-like finite-time touching rather than the capillary-dominated
long time thinning that would result in the absence of electric fields - this happens for $R=200$ and all values of $Q$ investigated here.
For order one values of $R$ the situation is more complex and non-monotonic
as evidenced by the phase diagram, but if $Q=\epsilon_1/\epsilon_2$
is not too large, then capillary instability dominates and inner wall touching derives.
 
The final computed shapes corresponding to $Q=1$ and the five different values of $R=0.002$, $0.02$, $2$, $20$, $200$ are
given in panels (a) to (f), respectively, in Fig.~\ref{dce}. An inner wall spike touching solution for the smallest value
$R=0.002$ gives way to
an outer wall spike touching solution for the largest value $R=200$, as explained physically above. 
In between we obtain inner wall touching solutions at large times
for moderate $R=0.02,\,0.2$, and outer wall touching ones for higher $R=2,\,20$.
A close inspection of the results reveals that inner wall touching solutions are accompanied by axial intervals where the interface is 
relatively flat near the outer wall - see for example panels (b) and (c) corresponding to $R=0.02$ and $0.2$, respectively.
The converse is true for outer wall touching, namely that it is accompanied by flat interface regions in the vicinity of the inner wall
as can be seen by the results in panels (d) and (e) for $R=2$ and $20$, respectively.
Regarding these local flat plateaus, we note that analogous patterns were reported by \cite{CM05} in their two-dimensional
study of electrified leaky dielectric liquid layers at zero Reynolds numbers confined between two parallel plates.
This is not surprising since our model can recover the equations of \cite{CM05}
in the limit of large inner tube radius, as was discussed at the end of section \ref{lweqns}.


\begin{figure}
  \centerline{ \includegraphics[width=6.in,height=4.in]{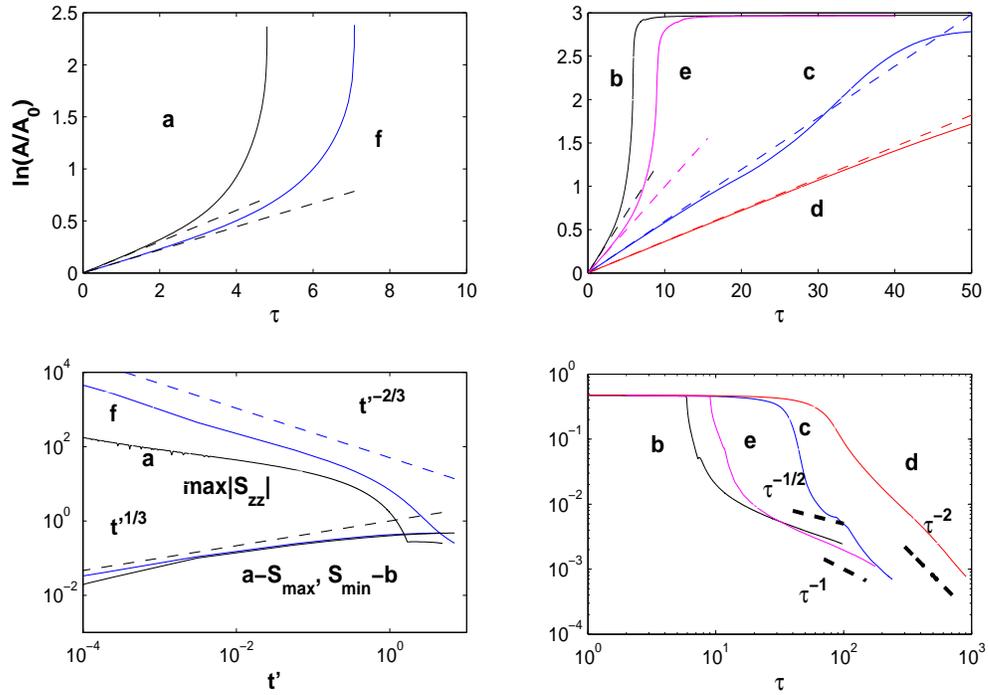} }
   \caption{Details of the simulation results corresponding to the insets in Fig.~\ref{dce}. 
   Top panels - evolution of scaled amplitude; bottom panels - evolution of minimum film thicknesses.
   The top and bottom left - cases (a) and (f) where finite-time touching solutions are obtained ($t'$ in the  bottom left 
   panel is the remaining time to the singularity). Top and bottom right - cases  (b)-(e) where film draining solutions are obtained; 
   solid lines represent for vanishing layer thickness $(S_{min}-b)$ or $(a-S_{max})$ as labelled. The dashed lines represent the 
   scaling behaviour predicted by the local analysis.}
   \label{dce1}
\end{figure}

More details of the numerical solutions depicted in Fig.\ref{dce} for $Q=1$ are considered next, and in particular we
describe the time evolution of solutions that lead to the terminal states depicted in the accompanying panels (a)-(f).
Results are given in Fig. \ref{dce1} - the left panels (top and bottom left) show the evolution for cases (a) and (f), while the
right panels correspond to panels (b)-(e). It can be seen from the top left panel that the evolution of the scaled interfacial amplitude $A/A_0$ (where
$A_0$ is the initial amplitude) for cases (a) and (f) terminates in finite time due to the interface hitting the upper or lower wall
with diverging speed. For cases (b)-(e) touching solutions develop at large times as confirmed by the saturation of $A/A_0$ in the
top right panel. In all instances the initial stages are in complete agreement with the predictions of the superimposed linear theory.

A detailed analysis of the data leading to the singular solutions for cases (a) and (f) is provided in the bottom left panel
of Fig.~\ref{dce1}. The evolution of the minimum film thickness (i.e. $a-S_{max}$ or $b-S_{min}$ for cases (a) and (f), respectively)
is shown as a function of $t^\prime =t_s-t$ where $t_s$ is the singular time (estimated in the computations), 
and it is seen to follow a scaling proportional to $(t^\prime)^{1/3}$. Further more the curvature of the interface $S_{zz}$ near the
rupture point is found to blow up in finite time with a scaling $(t^\prime)^{-2/3}$ as depicted in the results.
Both of these scalings are analogous with the perfectly conducting fluid results of \cite{WP2011}, and as discussed earlier this
is to be expected in the limits $R\ll 1$ and $R\gg 1$. Mathematically these limiting cases and their rupture behaviour can be anticipated
by considering $R\rightarrow 0$ to provide an equation similar to \eqref{spikefilmeqn}, while for $R\gg 1$ equation
(\ref{fen}) gives $E_N\sim -\frac{E_bQ}{2}\frac{1}{S^2\ln^2(S/b)}$. Therefore if $Q$ is away from zero, a finite time singular solution is expected.


Finally we consider the touching solutions found for the typical cases (b)-(e) when $R$ is not too small or too large.
The results summarised in the bottom right panel of Fig.~\ref{dce1} show a combination of the touching scalings that
have been described earlier. For case (b) that has $R=0.02$ we attain inner wall touching with the film draining as $\tau^{-1/2}$ for large $\tau$,
while for case (e) with $R=20$ the results suggest film thinning at a rate proportional to $\tau^{-1}$. These scalings are
in line with the Hammond solutions computed at large times by \cite{LRKCJ2006} and corresponding to lobe-collar and collar-collar solutions, respectively.
This behaviour can be understood further by analysing equation \eqref{sm1} in the vicinity of the inner wall $r=b$ or outer wall $r=a$, to
find that the leading order evolution is equivalent to that of the Hammond equation \eqref{hameqn}.
We can conclude, therefore, that the electric field does not modify the local behaviour of 
collar and lobe formation for $R$ relatively small or large.
Furthermore, by investigating the interfacial curvature we found that the maximum 
of $|S_{zz}|$ occurs in the vicinity of the corners of the flat plateau interfacial intervals 
that form on the wall opposite to the one where touching takes place (i.e. the outer wall for case (b) and the inner one for case (e)).
This behaviour is reminiscent of the quasi-static locally rectangular interfacial shapes computed by
\cite{CM05} in their two-dimensional study in channels.

However, we numerically observe a new scaling for the cases $(c)$ and $(d)$ corresponding to $R=0.2$ and $R=2$, for which
touching solutions at the inner and outer cylindrical walls are supported. In fact the results in Fig.~\ref{dce1} indicate that
for $R=0.2$ (case (c)) the vanishing film thickness at large times follows the scaling $a-S_{max}\sim \tau^{-2}$, while for $R=2$ (case (d))
the scaling is $S_{min}-a\sim \tau^{-2}$ -
these numerical results are shown in the  bottom right panel of Fig.~\ref{dce1} 
together with dashed lines to show the guiding scalings). 
We also find (in contrast to cases (b) and (e)) that the maximum of $|S_{zz}|$ occurs in the vicinity of 
the thinning region while the flat plateau regions remain away from wall and have relatively smooth edges in comparison to cases (b) and (e). 
In particular our numerical results indicate that $\delta^2 S_{zz}\sim (S_{min}-b)^{-1/4}$ in case (c) and $\delta^2 S_{zz}\sim O(1)$ in 
case (d) (data not shown). This leads us to suspect that the different scalings for (c) and (d) are transient and should
eventually behave as in cases (b) or (e). 
More detailed computations using adaptive mesh refinement appear to be necessary along with integrations to very large times,
and this is left for future work.

\section{Conclusions}\label{concl}
\setcounter{equation}{0}
We have investigated the linear and nonlinear dynamics of electrified two-fluid rod-annular flows. 
A novel evolution equation for the interface between the inner and outer fluids has been derived in the
long wave limit. 
The equation is general enough to allow for fully nonlinear interfacial deflections that can touch either the inner or
outer cylindrical walls - the assumption of a thin film relative to the cylinder radius is removed.
Meanwhile, the linear growth rate obtained from the derived equation has been compared to that of the full axisymmetric problem 
and excellent agreement is obtained for most cases providing strong evidence for the validity of the model when the
wavelength of interfacial perturbations is long compared to the annulus thickness.

The model retains numerous physical effects and is in turn characterised by a number of dimensionless parameters. 
These are geometry specific parameters measuring the relative thickness of the undisturbed layers and the
slenderness ratio $\delta$; fluid specific parameters
including an inner to outer fluid viscosity ratio $m$, a density ratio $\chi$ and a Reynolds number $R_e$ based on capillary scales;
electric field specific parameters including the permittivity ratio $Q$, conductivity ratio $R$ and an inverse electric Weber number $E_b$
that measures the strength of the applied electric field. In the present study we concentrate on electric field effects and so this large
parameter space is reduced by fixing $m=1$ and setting the undisturbed interface in the middle of the annulus (computations when the
undisturbed interface is near one of the walls were also carried out and discussed briefly). In addition the background flow is characterised
by small $R_e$ and our equations are inertialess.

Depending on the parameters a host of different nonlinear phenomena emerge. 
The non-electrified problem was studied first to establish that the interface evolves nonlinearly towards the vicinity of both the
outer and inner electrodes (at least for the range $1.3\lesssim c\lesssim 1.8$) where it forms thin films whose thicknesses decrease 
asymptotically at algebraic rates for large times. These results
are analogous to the lobe and collar structures described in detail for single fluid problems by \cite{Hammond1983} and \cite{LRKCJ2006}.
In addition it is found that if the undisturbed interface is sufficiently close to the outer or inner walls (in fact $c\gtrsim 1.8$ and $c\lesssim 1.3$,
respectively), then as expected one-sided touching takes place with the closest wall.

When electric fields are present a richer set of phenomena emerges depending on the parameters chosen in the leaky dielectric model.
These include finite-time singularities where the
interface touches the outer or inner wall with a locally cusp-like shape - for example for a fixed value
of the permittivity ratio $Q$, outer wall touching happens for very small
conductivity ratios $R$ and inner wall touching occurs for large $R$ (recall that $R=\sigma_2/\sigma_1$ is the outer to inner conductivity ratio).
At moderate values of $R$, however, one-sided asymptotic film thinning occurs on the inner wall for smaller
values of $R$ (e.g. for $Q=1$ and $R=0.02, \,0.2$), while for larger $R$ (e.g. $R=2,\,20$) the asymptotic film thinning structures
switch over to the outer wall. (Such phenomena are summarised collectively in Fig.~\ref{dce}.) 
Where asymptotic thinning takes place, our simulations indicate that the single-phase results based on the Hammond equation
(\citet{Hammond1983,LRKCJ2006}) are seen at least for relatively small or large values of the conductivity ratio $R$.
At moderate values of $R$, however, where one-sided thinning takes place we observe new scaling laws (see Fig.~\ref{dce1})
but we believe that these are transient in the sense that much longer time integrations are expected to give way to the Hammond
scalings accurately calculated by \cite{LRKCJ2006}. Our computations need to be altered significantly to accurately probe such regimes
and this is left for future work.
It is also found that in such cases of one-sided touching, portions of the interface become
relatively flat in the vicinity of the wall where asymptotic touching does not take place. Our simulations indicate that such structures are
associated with increasing values of the curvature where flat regions are connected with interfacial depressions or elevations.
Such phenomena where observed qualitatively by \cite{CM05} in their study of the two-dimensional problem and so we can conclude
that the dimension is not critical in the overall observed structures.

We also found that for perfect dielectric liquid pairs, the phenomena described above for the leaky dielectric models are also supported.
For example, starting with an undisturbed interface position $c=1.5$, large values of $Q=50$ predict cusp-like finite time touching singularities
on the outer wall. For moderate values of $Q$ (e.g. $Q=5,\,0.5$) asymptotic thinning takes place simultaneously (albeit at different rates),
while at smaller values of $Q$ (e.g. $Q=0.05$) one sided touchdown takes place on the inner wall with very intricate small scale drop structures
forming - see Figures \ref{pd1} and \ref{pd2}.

\section*{Acknowledgment}

The work of D.T.P. was supported in part by the Engineering and Physical Sciences Research Council of United Kingdom
by grants EP/K041134/1 and EP/L020564/1.

\begin{appendix}
\section{Stability matrix of long wave model}\label{lwM}
\setcounter{equation}{0}
In this section we record the components of the stability matrix derived from the linearised long wave model (\ref{la}), (\ref{lp}) and (\ref{ls}) in section \ref{linr}.
\[M= \begin{pmatrix}
  M_{11} & M_{12} &  M_{13}\\
 M_{21} & M_{22} & M_{23} \\
 M_{31} & M_{32} &  M_{33}
 \end{pmatrix}
\]
where 
\begin{align}
&M_{11}= ik\left(4G(c,b)E'_n+4c\ln(c/b)E'_t\right),\\
&M_{12}= ik\left(mF(c,a)-F(c,b)\right),\\
&M_{13}= -4\left(\ln(c/b)-m\ln(c/a)\right),\\
&M_{21}= -k^2\left(\left(\frac{F^2(c,b)}{16}-\frac{c^2G(c,b)}{2}\right)E'_n-cG(c,b)E'_t\right),\\
&M_{22}=\frac{mF^2(c,a)-F^2(c,b)}{16}k^2,\\
&M_{23}=ik(G(c,b)-mG(c,a)),\\
&M_{31}= c\omega, M_{32}=-\frac{F^2(c,a)}{16}k^2,M_{33}=ikG(c,a)
\end{align}
with $E'_n$ and $E'_t$ given by
\begin{align}
&E'_n = -\frac{1}{c^2} +\delta^2k^2 + E_b(1-QR^2)\frac{1+\frac{1-R}{\ln(c/a)-R\ln(c/b)}}{c^3\left(\ln(c/a)-R\ln(c/b)\right)^2}, \\
&E'_t= E_b(QR-1)R\frac{\ln(b/a)}{c^2\left(\ln(c/a)-R\ln(c/b)\right)^3}.
\end{align}
\end{appendix}

\section{Axisymmetric stability matrix for Stokes flow}\label{fullM}
\setcounter{equation}{0}
In this Appendix we consider the linear stability of the full axisymmetric problem at low Reynolds numbers. 
We consider the stability of the quiescent perfectly cylindrical steady state (bars are used to denote base-flow quantities):
\begin{align}
&\overline{\bm{u}}_j=\bm{0},\quad \overline{S}=c,\quad \left[\overline{p}\right]=\frac{1}{c}- \frac{E_b(1-QR^2)}{2}\frac{1}{c^2\left(R\ln(c/b)-\ln(c/a)\right)^2},
\nonumber\\
&\overline{\phi}_2=\frac{\ln(r/a)}{R\ln(c/b)-\ln(c/a)},\quad \overline{\phi}_1=\frac{R\ln(r/b)}{R\ln(c/b)-\ln(c/a)}+1\label{steadyB}
\end{align}
where $c$ denotes the position of the steady layer interface, $b<c<a$.
Perturbing about this state, linearising the boundary conditions and assuming normal mode solutions proportional
to $\exp(i\tilde{k}z+\omega t)$ allows for the problem to be cast into a rather simple eigenvalue problem for $\omega(\tilde{k})$,
with instability present whenever ${Real}(\omega)>0$. The perturbed interface position is described as
\begin{equation}
r=S(z,t)=c+\epsilon \hat{\eta}\exp(i\tilde{k}z+\omega t),
\end{equation}
where $\epsilon$ is a infinitesimally small. 
The flow is expressed in terms of a streamfunction so that $u_j=-(1/r)\partial\psi_j/\partial z$,
$w_j=(1/r)\partial\psi_j/\partial r$ and writing $\psi_j(r,z,t)=\widehat{\psi}(r)\exp(i\tilde{k}z+\omega t)$ provides the equation 
$E^4\widehat{\psi}_j=0$, where
the operator $E^2=\frac{d^2}{dr^2}-\frac{1}{r}\frac{d}{dr}-\tilde{k}^2$. The solutions are
\begin{eqnarray}
 \widehat{\psi}_j(r)=r\left(C_{1j}\,I_1(\tilde{k}r)+D_{1j}\,K_1(\tilde{k}r)+
 C_{2j}\,r\,I_{0}(\tilde{k}r)+D_{2j}\,r\,K_{0}(\tilde{k}r)\right),\label{psisoln}
\end{eqnarray}
where $C_{1j}$, $D_{1j}$, $C_{2j}$, $D_{2j}$ are unknown coefficients, and
$I_0, K_0$ are the modified Bessel functions of the first and second kind. 

The electric field is perturbed according to $\phi_j\sim \overline{\phi}_j+\epsilon \tilde{\phi}_j$, and
the solution for the linearised perturbation potential $\tilde{\phi}_j(r,z,t)=\widehat{\phi}_j(r)\,\exp(ikz+\omega t)$ proceeds along similar lines and satisfies zero order modified Bessel differential equation, 
\begin{equation}
\widehat{\phi}_j=A_j \left(I_0(\tilde{k}r)\,K_0(\tilde{k} d_j) - I_0(\tilde{k} d_i) \,K_0(\tilde{k}r)\right),\quad j=1,2
\end{equation} 
where $d_1=b$ and $d_2=a$.
Using the boundary conditions along the interface gives
\begin{eqnarray}
\frac{(R-1)\hat{\eta}}{c\left(R\ln(c/b)-\ln(c/a)\right)} + A_1\left( I_0(\tilde{k}c)K_0(\tilde{k}b) - I_0(\tilde{k}b) K_0(\tilde{k}c)\right)
=A_2\left( I_0(\tilde{k}c)K_0(\tilde{k}a) -I_0(\tilde{k}a) K_0(\tilde{k}c)\right),\end{eqnarray}
\begin{eqnarray}
A_1 \tilde{k}\left(I_1(\tilde{k}c)K_0(\tilde{k}b) + I_0(\tilde{k}b) K_1(\tilde{k}c)\right) = RA_2 \tilde{k}\left(I_1(\tilde{k}c)K_0(\tilde{k}a) + I_0(\tilde{k}a)K_1(\tilde{k}c) \right).
\end{eqnarray}
Turning to the fluid dynamics problem, at the tube wall we have the no slip and no penetration conditions,
\begin{equation}
\widehat{\psi}_1(r=b)=\hat{\psi}_2(r=a)=0,\qquad \widehat{\psi}^\prime_1(r=b)=\widehat{\psi}^\prime_2(r=a)=0. 
\end{equation}
Continuity of velocities at the fluid interface leads to 
\begin{equation}
\widehat{\psi}_1(r=c)=\widehat{\psi}_2(r=c),\quad \widehat{\psi}^\prime_1(r=c)=\widehat{\psi}^\prime_2(r=c). 
\end{equation}
Then the modified tangential and normal stress balances become
\begin{align}
&\frac{(1-m)}{c}\tilde{k}^2\widehat{\psi}_1-  \frac{1}{c^2}\widehat{\psi}^\prime_2
+ \frac{m}{c^2} \widehat{\psi}^\prime_1+\frac{1}{c}\widehat{\psi}^{\prime\prime}_2-\frac{m}{c}\widehat{\psi}^{\prime\prime}_1=
E_b i\tilde{k}\left(\widehat{\phi}_1+\widehat{\eta}\overline{\phi}_{1r}\right)  \left(QR-1\right)\overline{\phi}_{2r},\\
&\widehat{p}_2 -\widehat{p}_1 + \frac{2i\tilde{k}}{c^2}(m-1)\widehat{\psi}_1+\frac{2i\tilde{k}}{c}(\widehat{\psi}^\prime_2
-m\widehat{\psi}^\prime_1)=\frac{1}{c^2}\left(1-c^2\tilde{k}^2\right)\widehat{\eta}\nonumber\\
&+E_b\left(-\frac{\overline{\phi}_{2r}\widehat{\eta}}{c^2\left(R\ln(c/b)-\ln(c/a)\right)}+\overline{\phi}_{2r}\widehat{\phi}_{2r}
+Q\left(\frac{R\overline{\phi}_{1r}\widehat{\eta}}{c^2\left(R\ln(c/b)-\ln(c/a)\right)}-\overline{\phi}_{1r}\widehat{\phi}_{1r}\right)\right),
\end{align}
where $\widehat{\phi}_{j}$ is solved from the electrostatic problem and $\widehat{p}_j$ is obtained by inserting the velocities into the Stokes equations. Finally, the kinematic condition gives
\begin{equation}
c\,\omega\,\widehat{\eta}=-i\,\tilde{k}\,\widehat{\psi}_1.
\end{equation}

The above system needs to be solved for the eight unknowns ($C_{ij}$ and $D_{ij}$). After some computations we arrive at a linear and homogeneous system of 8 equations for the
unknown vector \begin{equation}\textbf{\emph{w}}=[C_{11},C_{21},C_{12},C_{22},D_{11},D_{12},D_{21},D_{22}]^T\end{equation} written in matrix form as
\begin{equation}
 \textbf{\emph{A}}\textbf{\emph{w}}=0.\label{eq:Mw}
\end{equation}
The coefficient matrix is given below (derivations in a similar context in the absence of electric fields
can be found in \cite{Pozrikidis2001a}, for example). For a non-trivial solution of (\ref{eq:Mw}) we require ${\rm{det}}(\bm{A})=0$ and this provides the desired dispersion relation that yields the growth-rate $\omega(\tilde{k})$. 
The relation to the long wave growth rate and wavenumber in our model is $\omega = \delta^2 \lambda$, $\tilde{k}=\delta k$, where $\delta$ is the slenderness parameter in the long wave model (this correspondence was used in our numerical comparisons between long wave
and exact linear theories in Section \ref{linr}).

The $64$ elements of the
matrix $\bm{A}=\left(a_{ij}\right)$ defined in equation (\ref{eq:Mw}) are given below for completeness.
\begin{align*}
a_{11}&=I_1(\tilde{k}c), \quad a_{12}=cI_0(\tilde{k}c),\quad a_{13}=K_1(\tilde{k}c),\quad a_{14}=cK_0(\tilde{k}c),\\ 
a_{15}&= -I_1(k),\quad a_{16}=-cI_0(\tilde{k}c), \quad a_{17}=-K_1(\tilde{k}c),\quad a_{18}=-cK_0(\tilde{k}c),\\
a_{21}&=\tilde{k}cI_0(\tilde{k}c),\quad a_{22}= 2I_0(\tilde{k}c) + \tilde{k}cI_1(\tilde{k}c),\quad a_{23}=-\tilde{k}cK_0(\tilde{k}c), \quad a_{24} = 2K_0(\tilde{k}c) - \tilde{k}cK_1(\tilde{k}c),\\
a_{25}&=-\tilde{k}cI_0(\tilde{k}c),\quad a_{26}= -2I_0(\tilde{k}c) - \tilde{k}cI_1(k),\quad a_{27}=\tilde{k}cK_0(\tilde{k}c),\quad a_{28}=-2K_0(\tilde{k}c)+\tilde{k}cK_1(\tilde{k}c)\\
a_{31}&=I_1(\tilde{k}b), \quad a_{32}=bI_0(\tilde{k}c), \quad a_{33}=K_1(\tilde{k}b), \quad a_{34}=bK_0(\tilde{k}b),\quad a_{35}=a_{36}=a_{37}=a_{38}=0,\\
a_{41}&=\tilde{k}I_0(\tilde{k}b),\quad a_{42}=2I_0(\tilde{k}b)+\tilde{k}bI_1(\tilde{k}b),\quad a_{43}=-kK_0(\tilde{k}b),\quad a_{44}=2K_0(\tilde{k}b)-\tilde{k}bK_1(\tilde{k}b)\\
 a_{45}&=a_{46}=a_{47}=a_{48}=a_{51}=a_{52}=a_{53}=a_{54}=a_{61}=a_{62}=a_{63}=a_{64}=0,\\
a_{55}&=   I_1(\tilde{k}a),\quad a_{56}=\tilde{k}aI_0(\tilde{k}a),\quad a_{57}=K_1(\tilde{k}a),\quad a_{58}=\tilde{k}aK_0(\tilde{k}a),\\
a_{65}&= kI_0(\tilde{k}a),\quad a_{66}=2I_0(\tilde{k}a)+\tilde{k}aI_1(\tilde{k}a),\quad a_{67}=-kK_0(\tilde{k}a),\quad a_{68}=2K_0(\tilde{k}a)-\tilde{k}aK_1(\tilde{k}a),
\end{align*}
\begin{align*}
a_{71}&= 2m kI_1(\tilde{k}c) - \mathcal{E}_c\, c I_1(\tilde{k}c),\quad a_{72} =2m (I_1(\tilde{k}c)+\tilde{k}cI_0(\tilde{k}c))-\mathcal{E}_c\, c^2I_0(\tilde{k}c),\\ 
a_{73}&=2m\tilde{k}K_1(\tilde{k}c)-\mathcal{E}_c\, cK_1(\tilde{k}c),\quad a_{74}=2m(-K_1(\tilde{k}c)+\tilde{k}cK_0(\tilde{k}c) - \mathcal{E}_c\, c^2 K_0(\tilde{k}c),\\
a_{75} &= -2\ k I_1(\tilde{k}c),\quad a_{76} = -2 (I_1(\tilde{k}c)+\tilde{k}cI_0(\tilde{k}c)),\quad a_{77}=-2kK_1(\tilde{k}c),\quad a_{78}=-2(\tilde{k}cK_0(\tilde{k}c)-K_1(\tilde{k}c)),\\
a_{81}&= \frac{2m}{c^2} \tilde{k}c (\tilde{k}cI_0(\tilde{k}c)-I_1(\tilde{k}c)) -\frac{\tilde{k}I_1(\tilde{k}c)}{c^2\omega}\left(1-c^2k^2+ E_bc^2(1-QR^2)\overline{\phi}_{2r}\left(\frac{ic\omega F_1}{R\,A_{1c}} - \frac{1}{c^2(R\ln(c/b)-\ln(c/a))}\right)\right),\\
a_{82}&= 2 m\tilde{k}^2c I_1(\tilde{k}c)-\frac{cI_0(\tilde{k}c)}{c^2\omega}\left(1-c^2k^2+ E_bc^2(1-QR^2)\overline{\phi}_{2r}\left(\frac{ic\omega F_1}{R\,A_{1c}} - \frac{1}{c^2(R\ln(c/b)-\ln(c/a))}\right)\right),\\
a_{83}&= -\frac{2m}{c^2} \tilde{k}c (\tilde{k}cK_0(\tilde{k}c)+K_1(\tilde{k}c)) -\frac{\tilde{k}K_1(\tilde{k}c)}{c^2\omega}\left(1-c^2k^2+ E_bc^2(1-QR^2)\overline{\phi}_{2r}\left(\frac{ic\omega F_1}{R\,A_{1c}} - \frac{1}{c^2(R\ln(c/b)-\ln(c/a))}\right)\right),\\
a_{84}&= -2 m\tilde{k}^2c K_1(\tilde{k}c)-\frac{cK_0(\tilde{k}c)}{c^2\omega}\left(1-c^2k^2+ E_bc^2(1-QR^2)\overline{\phi}_{2r}\left(\frac{ic\omega F_1}{R\,A_{1c}} - \frac{1}{c^2(R\ln(c/b)-\ln(c/a))}\right)\right),\\
a_{85}&= -2 \tilde{k}/c(\tilde{k}cI_0(\tilde{k}c)-I_1(\tilde{k}c)),\quad a_{86} = -2\tilde{k}^2c I_1(\tilde{k}c),\\
a_{87} &= 2 \tilde{k}/c(\tilde{k}cK_0(\tilde{k}c)+K_1(\tilde{k}c)),\quad a_{88} = 2 \tilde{k}^2c K_1(\tilde{k}c).
\end{align*}
where 
\begin{align}
F_0 &= I_0(\tilde{k}c)\,K_0(\tilde{k}b) - I_0(\tilde{k}b)\,K_0(\tilde{k}c),\\
F_1& = I_1(\tilde{k}c)\,K_0(\tilde{k}b) + I_0(\tilde{k}b)\,K_1(\tilde{k}c),\\
G_0 & = I_0(\tilde{k}c)\,K_0(\tilde{k}a) - I_0(\tilde{k}a)\,K_0(\tilde{k}c),\\
G_1& = I_1(\tilde{k}c)\,K_0(\tilde{k}a) + I_0(\tilde{k}a)\,K_1(\tilde{k}c),\\
A_{1c} & = \frac{-i\tilde{k}R(R-1)G_1}{(F_1G_0-RF_0G_1)(c^2\omega (R\ln(c/b)-\ln(c/a)))},\\
\mathcal{E}_c & = i\,E_b(QR-1)\left(A_{1c}\,F_0 - \frac{i\tilde{k}\,\overline{\phi}_{1r}\big{|}_{r=c}}{c\omega}\right)\overline{\phi}_{2r}\big{|}_{r=c}.
\end{align}

\bibliographystyle{imamat}
\bibliography{elayer}
%


\end{document}